\documentclass[fleqn,usenatbib]{mnras}
\usepackage[normalem]{ulem}
\usepackage{newtxtext,newtxmath}

\usepackage[T1]{fontenc}
\usepackage{ae,aecompl}

\usepackage{graphicx}	
\usepackage{amsmath}	
\usepackage{amssymb}	
\usepackage{amsfonts}
\usepackage{color}
\usepackage{float}
\usepackage{booktabs}
\usepackage{multirow}
\usepackage[english]{babel}
\usepackage[utf8]{inputenc}

\newcommand*\xbar[1]{%
   \hbox{%
     \vbox{%
       \hrule height 0.5pt 
       \kern0.5ex
       \hbox{%
         \kern-0.1em
         \ensuremath{#1}%
         \kern-0.1em
       }%
     }%
   }%
} 

\title[Destruction and multiple ionization of PAHs]{Destruction and multiple ionization of PAHs by X-rays in circumnuclear regions of AGNs}

\author[T. Monfredini et al.]{
Thiago Monfredini,$^{1,2}$
Heidy M. Quiti\'an-Lara,$^{1}$
Felipe Fantuzzi,$^{3,4}$
Wania Wolff,$^{5}$
\newauthor Edgar Mendoza$^{1,6}$
Alexsandre F. Lago,$^{7}$
Dinalva A. Sales,$^{8}$
Miriani G. Pastoriza,$^{9}$
\newauthor and Heloisa M. Boechat-Roberty$^{1}$\thanks{E-mail: heloisa@astro.ufrj.br}
\\
$^{1}$Observat\'orio do Valongo, Universidade Federal do Rio de Janeiro, Ladeira Pedro Ant\^onio, 43, Rio de Janeiro - RJ, \\ 20.080-090, Brazil \\
$^{2}$Laborat\'orio Nacional de Luz S\'incrotron, Centro Nacional de Pesquisas em Energia e Materiais, Polo II de Alta Tecnologia,\\
Campinas - SP, 13083-100, Brazil\\
$^{3}$Instituto de Qu\'imica, Universidade Federal do Rio de Janeiro, Cidade Universitária, Rio de Janeiro - RJ, 21941-909, Brazil\\
$^{4}$Current Address: Institute for Inorganic Chemistry, Julius-Maximilians-Universität Würzburg, Am Hubland, 97074 Würzburg, Germany.\\
$^{5}$Instituto de F\'isica,  Universidade Federal do Rio de Janeiro, Cidade Universitária, Rio de Janeiro - RJ, 21941-972, Brazil\\
$^{6}$Instituto de Astronomia, Geof\'isica e Ciências Atmosféricas, Universidade de S\~ao Paulo, Cidade Universitária, São Paulo - SP,\\ 
05508-090, Brazil\\
$^{7}$Centro de Ci\^encias Naturais e Humanas, Universidade Federal do ABC, Santa Terezinha, Santo Andr\'{e} - SP, 09210-580, Brazil\\
$^{8}$Instituto de Matem\'atica, Estat\'istica e F\'isica, Universidade Federal do Rio Grande, Carreiros, Rio Grande - RS, 96201-900, Brazil\\
$^{9}$Instituto de F\'isica, Universidade Federal do Rio Grande do Sul, Agronomia, Porto Alegre - RS, 91501-970, Brazil\\
}

\date{Accepted XXX. Received YYY; in original form ZZZ}

\pubyear{2018}

\begin{document}
\label{firstpage}
\pagerange{\pageref{firstpage}--\pageref{lastpage}}
\maketitle

\begin{abstract}
The infrared signatures of polycyclic aromatic hydrocarbons (PAHs) are observed in a variety of astrophysical objects, including the circumnuclear medium of active galactic nuclei (AGNs). These are sources of highly energetic photons (0.2 to 10 keV), exposing the PAHs to a harsh environment. In this work, we examined experimentally the photoionization and photostability of naphthalene (C$_{10}$H$_{8}$), anthracene (C$_{14}$H$_{10}$), 2-methyl-anthracene (C$_{14}$H$_{9}$CH$_{3}$) and pyrene (C$_{16}$H$_{10}$) upon interaction with photons of 275, 310 and 2500 eV. The measurements were performed at the Brazilian Synchrotron Light Laboratory using time-of-flight mass-spectrometry (TOF-MS). We determined the absolute photoionization and photodissociation cross sections as a function of the incident photon energy; the production rates of singly, doubly and triply charged ions; and the molecular half-lives in regions surrounding AGNs. Even considering moderate X-ray optical depth values ($\tau = 4.45$) due to attenuation by the dusty torus, the half-lives are not long enough to account for PAH detection. Our results suggest that a more sophisticated interplay between PAHs and dust grains should be present in order to circumvent molecular destruction. We could not see any significant difference in the half-life values by increasing the size of the PAH carbon backbone, N$_C$, from 10 to 16. However, we show that the multiple photoionization rates are significantly greater than the single ones, irrespective of the AGN source. We suggest that an enrichment of multiply charged ions caused by X-rays can occur in AGNs.
\end{abstract}




\section{Introduction}

Neutral and ionized Polycyclic Aromatic Hydrocarbons (PAHs) are detected in astronomical sources through emission bands in the infrared (IR) wavelength range, due to the corresponding molecular vibrations. \citep{Peeters2002a,Tielens2008}. The main vibrational features of PAHs comprise the \-{C$-$H} (3.3 $\mu$m), \-{C$=$C} (6.2 $\mu$m) and \-{C$-$C} (7.7 $\mu$m) bond stretching modes, as well as the in-plane (8.6 $\mu$m) and out-of-plane (11.3 $\mu$m) \-{C$-$H} bond bending modes. From the analysis of such IR bands, PAHs have been observed in a diversity of galactic objects, such as planetary nebulae \citep{Waters1998,Gorny2001,Ohsawa2012,Guzman-Ramirez2014}, HII regions \citep{Roelfsema1996,Peeters2002}, reflection nebulae \citep{Boersma2014,Ricca2018}, protoplanetary disks \citep{Ressler2003,Maaskant2014,Schworer2017,Seok2017,Taha2018}, among others. The PAH IR features contribute to about 10\% of the ISM luminosity in the 1-1000 $\mu$m range, accounting for a large fraction of the elemental C in star-forming galaxies \citep{Lagache2004}. In addition, their luminosities are well correlated with star formation rates \citep{Peeters2004,Stierwalt2014,Alonso-Herrero2014,Esparza-Arredondo2018}. In view of such remarkable characteristics, continuing research on the formation and stability of PAHs has a key relevance to astrochemistry.

In addition to galactic sources, PAHs have also been observed in a variety of extragalactic objects, such as HII regions in the Magellanic Clouds \citep{Li2002a,Vermeij2002,Oey2017}, local dusty elliptical galaxies \citep{Kaneda2008}, starburst galaxies \citep{Brandl2006}, submillimeter galaxies (SMGs) \citep{Menendez-Delmestre2009}, ultra-luminous IR galaxies (ULIRGs) \citep{Desai2007} and in the circumnuclear regions of Active Galactic Nuclei (AGNs) \citep{Lutz1998,ODowd2009,Tommasin2010,Sales2013,Esquej2013,Alonso-Herrero2016}. Concerning the latter objects, PAHs have been identified in both Seyfert 1 and Seyfert 2 galaxies \citep{Mazzarella1994,Deo2007, Diamond-Stanic2010, Sales2013}, in low-ionization nuclear emission-line regions (LINERs) \citep{Sturm2006} and obscured quasars \citep{MartinezSansigre2008}. 

AGNs are sources of both soft (0.2-2 keV) and tender (2-10 keV) X-ray radiation, and contribute significantly to the extragalactic X-ray background \citep{Comastri1994,Mushotzky2000,Lubinski2016,Hickox2018}. The X-ray emission mechanism is known to be powered by gas accretion on to a central supermassive black hole \citep{Ferrarese2005,Zhang2005,DiMatteo2005,Barai2012}, which is the basis of the standard unification model for AGNs \citep{Antonucci1993,Urry1995} --- for a recent review, see \citealt{Netzer2015}. The combination of a 0.2-10 keV radiation field and the presence of PAHs in the AGN vicinity provides an interesting scenario in which laboratory investigation of molecular photoionization and photodestruction in the X-ray energy range could provide important information. 

The  experimental studies of vacuum ultraviolet (VUV) photoionization and photodissociation of PAHs were firstly reported at about thirty years ago by \cite{Leach1989, Leach1989b}, using time-of-flight mass-spectrometry and photoelectron-photoion coincidence techniques. \cite{Jochims1994,Jochims1996,Jochims1999} have also investigated the VUV photostability of PAHs and methyl-substituted PAHs. In addition,
the competition between VUV photoionization, photofragmentation and the photoproduction of dications in the context of interstellar PAH population were discussed by \citealt{Zhen2016}. Using soft X-ray photons with energies around the C1s$\rightarrow \pi$*  resonance (285 eV), \cite{Reitsma2014,Reitsma2015} have investigated the fragmentation of PAHs cations. However, from our perspective, there is still a lack of information on the ionization and dissociation of PAHs driven by photons of higher energies, which would be useful in the context of the circumnuclear regions of AGNs. The photochemistry of these regions should be highly affected by a radiation field with 0.2-10 keV photons due to its high penetration power, even through a gas with column densities up to 10$^{24}$ cm$^{-2}$, such as X-ray dominated regions (XRD, see \citealt{Usero2004}) .

 \begin{figure}
 \centering
      \includegraphics[width=0.7\columnwidth]{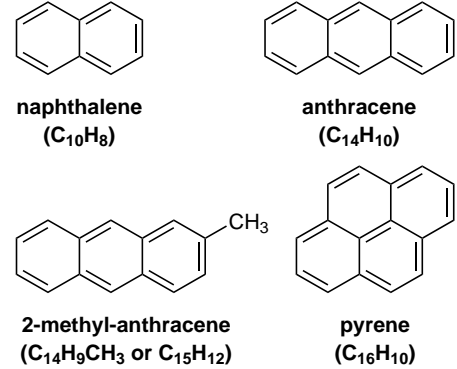}    
     \caption{The polycyclic aromatic hydrocarbons studied in this work.}
     \label{fig:1molecules}
 \end{figure}

The present work aims at studying the effect of X-ray photon interactions on the photodissociation and photoionization rates of the PAH molecules shown in Fig. \ref{fig:1molecules}. In order to compare the ionic fragmentation profiles with respect to the photon energy, we have measured mass spectra at selected energies below (275 eV) and above (310 eV) the C1s resonance features of these PAHs, and at 2500 eV. From the mass spectra obtained herein, we determined the photoionization and photodissociation cross-sections. These results are discussed in the context of the chemistry of the circumnuclear regions of AGNs, which is the focus of our study.

\section{Experimental Setup}
\label{sec:exp}

The measurements of PAH molecules fragmentation were performed using X-ray photons from two different experimental beamlines of the the Brazilian Synchrotron Light Laboratory at the National Center for Energy and Materials Research (LNLS/CNPEM), Campinas, São Paulo. For photon energies around the C1s core level (275 eV and 310 eV) we used the Spherical Grating Monochromator (SGM), which covers the energy range from 250 to 1000 eV. For 2500 eV, we used the Soft X-ray Spectroscopy (SXS) beamline, which operates in the 1000-5000 eV energy range. The selection of the specific photon energies covered in this work were also based on the fact that they are among the highest photon fluxes in each beamline, which ultimately results in better ion counting and low signal-to-noise ratio. For both beamlines,  the photon flux at the sample was around 10$^{11}$ photons s$^{-1}$mm$^{-2}$, with an energy resolving power of $E / \Delta E > 3000$ \citep{Fonseca1992, Abbate1999, Wolff2012}.

The experimental setup for the SGM and SXS beamlines has been described in detail elsewhere \citep{Lago2004,Boechat2009,Lago2017}. Briefly, X-ray photons perpendicularly intersect the gas sample inside a ultra-high vacuum chamber ($\lesssim$ 10$^{-8}$ mbar), leading to molecular ionization and dissociation, from which the ion products were analyzed and studied. 

A Wiley and McLaren time-of-flight (TOF) spectrometer was employed for the charge and mass analysis of the ionic species \citep{Burmeister2010}. The ionized species produced by the interaction of the sample with the photon beam were accelerated and focused by a two-stage electric field, traveled through a 297 mm field free drift tube and detected by a pair of microchannel plate (MCP) detectors mounted in a chevron configuration. The high voltage of \mbox{-4250 V} applied on the front of a MCP detector assured the independence of the efficiencies of the ions on the ion mass and charge state. The photoelectrons, accelerated in the opposite direction with respect to the positive ions, focused by an electrostatic lens set, were recorded by two MCP also arranged in a chevron configuration. A 500 V/cm DC electric field was applied to the first ion acceleration stage.  Electrons with initial kinetic energies up to 150 eV, and ions up to 30 eV were detected without energy analysis. 
During the experiments, PhotoElectron PhotoIon (PEPICO) and PhotoElectron PhotoIon PhotoIon Coincidence (PEPIPICO) spectra, where at least one of the ejected electrons and one or two positive ions are recorded in coincidence, were collected. In this work, we only report the study of the coincidences between one electron and one cation (PEPICO), at selected photon energies, in order to identify the positive ions resulting from the photoionization process and their relative intensities. The electronics setup allows for simultaneous multi-hit capability within 1 ns resolution. The start (photoelectron) and stop (photoion) signals were then correlated, converted from the time domain (nanosecond range) to the mass/charge ratio (m/z) domain, and finally expressed in terms of intensity as a function of m/z, resulting in PEPICO mass spectra. The number of single and double coincidence outcomes were corrected adopting the routines of \cite{Simon1991}, where the electron and ion detection efficiencies were taken into account. The calibration of the TOF spectrometer was accomplished by identifying the peaks related to the different samples. 
All PAH samples, namely naphthalene (C$_{10}$H$_{8}$), anthracene (C$_{14}$H$_{10}$), 2-methyl-anthracene (C$_{14}$H$_{9}$CH$_{3}$, or C$_{15}$H$_{12}$) and pyrene (C$_{16}$H$_{10}$), whose structures are shown in Fig. \ref{fig:1molecules}, were purchased from SIGMA-Aldrich. They were obtained in the dry powder form with purities higher than 99.5\% and used without further purification. In this work, we studied the PAH molecules in the gas phase and a heated source for sublimation was used. A schematic representation of the sublimation setup is described in detail in \cite{Lago2004}. It basically consists of a small stainless-steel cylinder designed to store a maximum of 800 mg of the solid sample, involved by an isolated low resistance tungsten wire, a controllable DC low voltage source, and a thermocouple. In order to control the output flux of the sample vapor we employed thin layers of inert quartz meshes between sample powder portions. The samples were slowly heated by increasing the DC voltage applied to the wire, in steps, while monitoring the sample temperature and the experimental chamber pressure, during the necessary time to reach the optimal range of sublimation temperatures for the experiments. At least one hour was usually necessary for the sublimation temperature to reach equilibrium. The working temperatures of the PAH samples were in the range of 115 to 165$^{o}$C. The experiment was conducted usually 20–30$^{o}$C above the temperature at which a molecular jet was firstly noted. Information on the sublimation of PAHs, as well as on the extent of decomposition, is available elsewhere \citep{Welsh1997}. The sublimated sample was diffused through a heated nozzle of 0.8 mm of internal diameter, to avoid clogging. This heating apparatus was inserted into the vacuum chamber and the effusive gaseous jet was intercepted by the photon beam. The alignment of the tip of the needle was checked with the emergent photon beam recorded by a light sensitive diode. The pressure was kept in the range of 10$^{-6}$ mbar during experiments to avoid charge transfer between the ions and the background and to assure the single photon/molecule collision regime.

\section{Results and Discussion}
\label{sec:results}

\subsection{Mass Spectra and Production of Ions} \label{sub:mass}

The PEPICO mass spectra were obtained by following the data treatment methodology described by \cite{Boechat2005}. Briefly, after subtraction of the background noise and false coincidences coming from aborted double and triple ionizations \citep{Simon1991}, a standard deconvolution method is used to determine the Partial Ion Yield -- $PIY$ -- of each fragment-ion, defined as:

\begin{equation}
PIY = 100  \left (\dfrac{A_{i}}{\sum A_{i}} \right),
\end{equation} where $A_i$ is the area of the peak of each ion $i$, $\sum A_i$ is the sum of the areas of all peaks in the spectrum and PIY is given in percentage. The $A_i$ values were corrected by considering the ion and electron detection efficiencies obtained by the work of \cite{Burmeister2010}, for this same TOF-MS and detectors setup. The uncertainties of the partial yields are estimated around 10\% and 15\% for the highest and lowest peak areas, respectively. The uncertainty in the peak position reaches m/z of 0.5. In conjunction to the singly and multiply charged parent ions, we observe the appearance of peaks due to the carbon isotope $^{13}$C, whose intensities are in agreement with the natural isotopic abundance (1.11\%). In the cases where the m/z ratio of doubly charged parent molecule coincides with a singly charged fragment, we could discriminate the contribution of both species by knowing the natural abundance of the doubly ionized isotopologue and determining its area. It was possible to discriminate two different peak profiles depending on the charge state: the full width at half maximum (FWHM) of singly charged ions are broader than those of the dicationic species. The $PIY$s of the fragment-ions identified in this work are listed in the Appendix, for all measured photon energies.

\begin{figure*}
\centering
  \includegraphics[width=\textwidth]{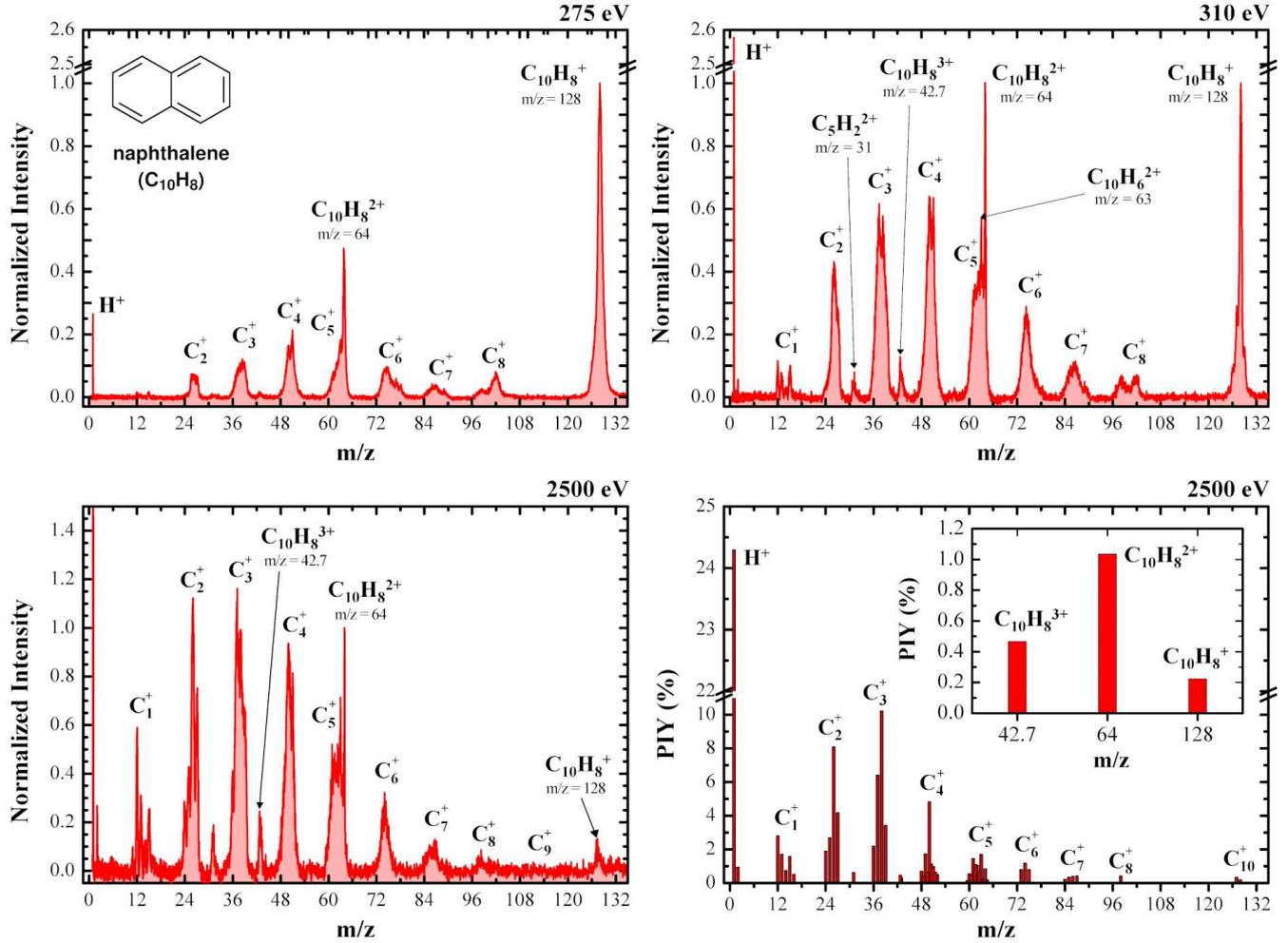}
    \caption{Time of flight mass spectra of ionic fragments due to the X-ray photodissociation of naphthalene at energies of 275 eV (top left), 310 eV (top right) and 2500 eV (bottom left). Bottom right: Partial ion yield, $PIY$(\%), recorded at 2500 eV with an inserted figure highlighting the production of singly, doubly and triply charged parent ions of C$_{10}$H$_8$.}
    \label{fig:2naph-spec}
\end{figure*}

Figure \ref{fig:2naph-spec} shows the mass spectra of naphthalene triggered by X-ray photons of 275 eV, 310 eV and 2500 eV. For 275 eV and 310 eV the plot shows the normalized intensity with respect to the counts of the parent ion. On the other hand, for 2500 eV the intensity values are normalized with respect to the counts of the doubly charged parent ion. For 275 eV, in which the photon energy is below the C1s resonance energy (285 eV), the spectrum is dominated by the C$_{10}$H$_8^+$ (m/z $ = $ 128) parent ion. The strong feature at m/z $=$ 64, with normalized intensity of $\sim$0.5, corresponds to the doubly charged parent ion, C$_{10}$H$_8^{2+}$. Such species is expected to play an important role in the interstellar chemistry, since its small appearance energy suggests that it could be formed in diffuse HI regions \citep{Malloci2007}. In addition, the C$_{10}$H$_8$ dication is the most abundant product from the reaction between He$^+$ and naphthalene at thermal energies \citep{Petrie1993}, which also has implications for astrochemistry. To the best of our knowledge, there is still no consensus on the global minimum structure of C$_{10}$H$_8^{2+}$. The planar aromatic 10-membered ring depicted by \cite{Leach1989} and the bicyclic structure originated by the junction of the pentagonal pyramidal benzene dication \citep{Jasik2014,Fantuzzi2017a} with the neutral benzene hexagonal ring are among the most prominent candidates.

Apart from the singly and doubly charged parent ions, Figure \ref{fig:2naph-spec} (top panels) also presents features related to dissociation products in which the carbon backbone varies from groups of 2 up to 8 C atoms. However, for 275 eV, the normalized intensities of such groups are small in comparison with the singly charged parent ion peak. This indicates that the carbon backbone rupture is a secondary process for energies below the C1s resonance energy. On the other hand, this picture is drastically affected by the increase of the photon energy. For 310 eV, depicted in Figure \ref{fig:2naph-spec} (top right), the non-dissociative single ionization process is much less prominent than in the previous case, and the normalized intensities of the single and double ionization features are equivalent. In addition, the carbon backbone fragmentation processes are more competitive to ionization, especially the ones related to the formation of molecular ions bearing 3 (group C$_3^+$) and 4 (group C$_4^+$) carbon atoms. Another interesting aspect of this mass-spectrum is the presence of a prominent feature at m/z $=$ 42.7, which can be attributed to the triply ionized parent ion. The photo-electrons extracted from the core of the naphthalene molecule, therefore, lead to Auger processes that ultimately produce double and triple ionization. \cite{Reitsma2015} have also detected the production of di- and trications resulting from photoionization and photodissociation of free coronene cations (C$_{24}$H$_{12}^+$) upon interaction with soft X-rays photons of the C1s energy range (283-305 eV). {In addition, they have shown} that the relative production of trications overcomes the dication production for energies higher than 294 eV, which is not observed for naphthalene in our experiments. This is a clear indication that the increase of the PAH carbon backbone size could facilitate the formation of multiply charged species, as will be discussed in the next sections.

\begin{figure*}
\centering
       \includegraphics[width=16.5cm]{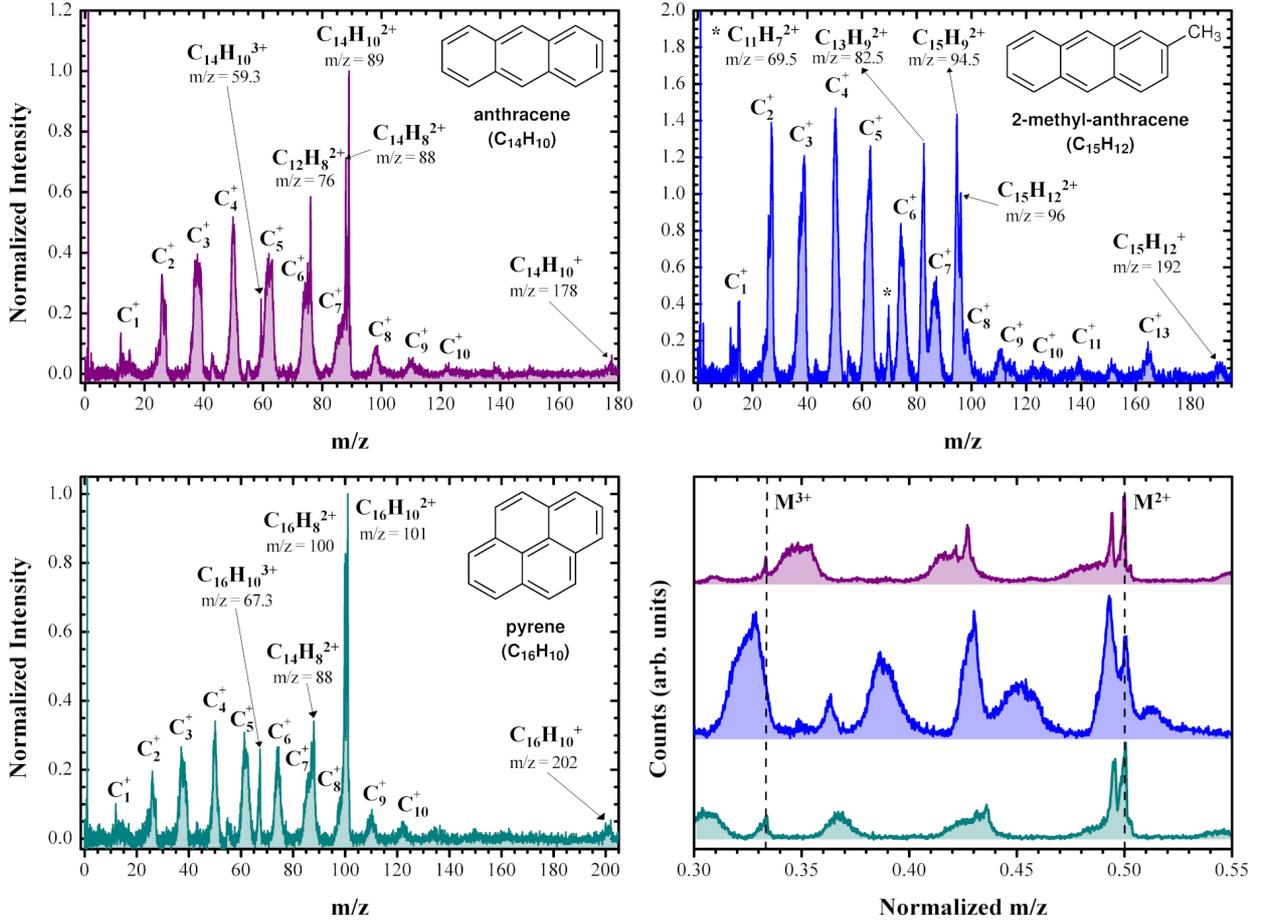}
    \caption{Time of flight mass spectra of ionic fragments due to the  X-ray photodissociation of anthracene (top left), 2-methyl-anthracene (top right) and pyrene (bottom left) at 2500 eV. Bottom right: The same mass spectra normalized by the mass of the parent molecule, in the range between 0.3 and 0.5 of the parent ion m/z.}
    \label{fig:4MS_anthr_meth_py}
\end{figure*}

The inner shell excitation process triggered by X-ray photoabsorption is initiated by the formation of a vacancy in  the C1s level, which is subsequently filled by the decay of a valence electron via radiative (X-ray emission) or via nonradiative (Auger) processes. The latter is usually a dominant process. It involves, in most cases, the ejection of an Auger electron to balance the energy excess of the excited ion state, which results in the production of multiply charged species. The Auger process causes molecular instabilities that can activate dissociation channels into neutral and ionic fragments. Ultimately, such molecular species could act as molecular seeds for the nucleation and growth of larger PAHs in astrophysical environments, following similar mechanisms such as the one experienced by the doubly charged benzene dimer \citep{Deleuze2005}.

Very little is known about the triply charged C$_{10}$H$_8$ species, apart from the fact that it is also observed in experiments using intense laser fields \citep{Ledingham1999}. A small feature is also observed in the spectrum at m/z $=$ 31.0, which could be attributed to C$_5$H$_2^{2+}$ (or less likely to C$_{10}$H$_4^{4+}$).  

Figure \ref{fig:2naph-spec} (bottom left) shows the mass spectrum of ionic fragments from naphthalene dissociation after interaction with a 2500 eV photon. In this case, it is noticeable that the yield of the single ionization process is severely affected in comparison with the previous cases, and the spectrum is dominated by the photofragmentation of the carbon backbone. The ion yield of the species with m/z values above half of the parent ion are remarkably low, possibly due to a Coulomb explosion effect \citep{Hoshina2011}. The features related to multiple ionization processes, on the other hand, increase from 275 eV to 2500 eV, as revealed by their $PIY$ values (see Table \ref{tab-all} in the Appendix).

Figure \ref{fig:4MS_anthr_meth_py} shows the mass spectra of anthracene (top left), 2-methyl-anthracene (top right) and pyrene (bottom left) at 2500 eV. The intensities are normalized with respect to the ones associated to the double ionization process. In the bottom right panel of Figure \ref{fig:4MS_anthr_meth_py}, we also show a comparison between the three plots by normalizing the m/z axis of each spectrum with respect to the m/z value of the respective parent ion, highlighting the positions of the double and triple ionization peaks. Although each mass spectrum in the figure has its own characteristics, some general trends can be observed. In all cases, the feature related to the single ionization species is significantly small in comparison to its double and triple ionization analogs. In addition, for anthracene and pyrene the shapes of the carbon backbone dissociation groups are quite similar, while the presence of the methyl group in 2-methyl-anthracene leads to a more distinct pattern. As indicated in the bottom right panel of Figure \ref{fig:4MS_anthr_meth_py}, we were not able to identify the triply ionized methyl-anthracene species, while the trication of anthracene and pyrene are unambiguously observed.  

For anthracene, besides the identification of C$_{14}$H$_{10}^{2+}$ (m/z $=$ 89), the C$_{14}$H$_{8}^{2+}$ (m/z $=$ 88) and C$_{12}$H$_{8}^{2+}$ (m/z $=$ 76) dication species are also easily observed. The C$_{14}$H$_{8}^{2+}$ species is originated by the loss of 2 H atoms from the doubly charged anthracene molecule, while the latter ion is formed after a C$_2$H$_2$ loss. In fact, these are among the main dissociation channels for the anthracene dication as attributed by previous works \citep{Holm2011,Reitsma2012}. The $PIY$(\%) values for C$_{14}$H$_{10}^{2+}$, C$_{14}$H$_{8}^{2+}$ and C$_{12}$H$_{8}^{2+}$ obtained herein are 1.1\%, 0.7\% and 1.0\%, respectively. The main carbon backbone photodissociation pathways, on the other hand, are the ones leading to the formation of C$_3$H$^+$ (6.4\%), C$_3$H$_3^+$ (6.0 \%), C$_2$H$_2^+$ (5.3 \%), C$_4$H$_2^+$ (4.8 \%) and C$_3$H$_2^+$ (4.4 \%). Our results show that the dissociation channels leading to C$_3$H$_m^+$ ions are more active than the ones leading to C$_2$H$_m^+$. The same dominance of C$_3^+$ over C$_2^+$ was observed by \cite{Reitsma2012} on the fragmentation of anthracene after a double electron transfer process to a 5 keV proton.

We have also identified the production of doubly charged species in the spectrum of 2-methyl-anthracene at 2500 eV. Besides the C$_{15}$H$_{12}^{2+}$ dication, which is resulted from the non-dissociative double photoionization of the parent molecule, the species C$_{15}$H$_{9}^{2+}$, C$_{13}$H$_{9}^{2+}$ and C$_{11}$H$_{7}^{2+}$, m/z = 94.5, 82.5 and 69.5 respectively, were easily discriminated in the mass spectrum. The C$_{15}$H$_{9}^{2+}$ ion is produced by the detachment of 3 H atoms from C$_{15}$H$_{12}^{2+}$, a type of dissociation which is pointed by \cite{Mathur1981} as a common feature in the doubly charged ion mass spectra of organic ring molecules containing an odd number of carbon atoms. On the other hand, the C$_{13}$H$_{9}^{2+}$ species is generated from the loss of C$_2$H$_2$ $+$ H. The fragment with m/z = 165 is attributed to the fluorenyl cation (C$_{13}$H$_{9}^{+}$), which presents a central five-membered antiaromatic ring with formal 4$\pi$ electrons \citep{Costa2015}. However, to the best of our knowledge, the doubly charged analog of the fluorenyl cation has not been studied in detail so far. Finally, the C$_{11}$H$_{7}^{2+}$ is produced from the loss of 2 C$_2$H$_2$ $+$ H. The singly charged C$_{11}$H$_{7}^{+}$ species is observed as one of the major secondary ions obtained from the reaction of benzene with its electron bombardment products \citep{Lifshitz1969}. The C$_{11}$H$_{7}^{2+}$ ion, on the other hand, was not described in the literature up to this work. The major carbon backbone photodissociation products of 2-methyl-anthracene are C$_{2}$H$_{3}^{+}$ (9.8\%), C$_3$H$_3^+$ (9.2\%), C$_3$H$^+$ (5.0\%), C$_2$H$_2^+$ (4.6\%) and C$_5$H$_3^+$ (4.1\%). 

For pyrene, in addition to the C$_{16}$H$_{10}^{2+}$ (m/z $=$ 101) dication, the doubly charged C$_{16}$H$_{8}^{2+}$ (m/z $=$ 100) and C$_{14}$H$_{8}^{2+}$ (m/z $=$ 88) species are also easily discriminated in the spectrum. The former is originated by the loss of 2 H from the parent dication, while the latter results from a C$_2$H$_2$ loss. Such dicationic species are also observed by the collision of pyrene with electrons \citep{Kim2001} and multiply charged ions \citep{Lawicki2011}.

The main C$_n^+$ ions produced by our experiment with pyrene are C$_2$H$_2^+$ (5.6\%), C$_{16}$H$_{8}^{2+}$ (4.9\%), C$_4$H$_2^+$ (4.9\%), C$_3$H$_3^+$ (3.4\%) and C$_3$H$^+$ (3.1\%). The mass spectrum of pyrene suggests that there is a very low contribution of singly charged ions with the same m/z ratio as doubly charged fragments. In these cases, with exception of the parent dication (C$_{16}$H$_{10}^{2+}$) and its isotopologue, the signal was evaluated as coming entirely from the doubly charged ions. The deconvolution signal processing can lead to overestimation as well as additional uncertainty to the $PIY$s. In any case, there is strong evidence that multiple ionization processes become more feasible as the PAH carbon backbone size increases.

\begin{figure*}
\centering
       \includegraphics[width=17cm]{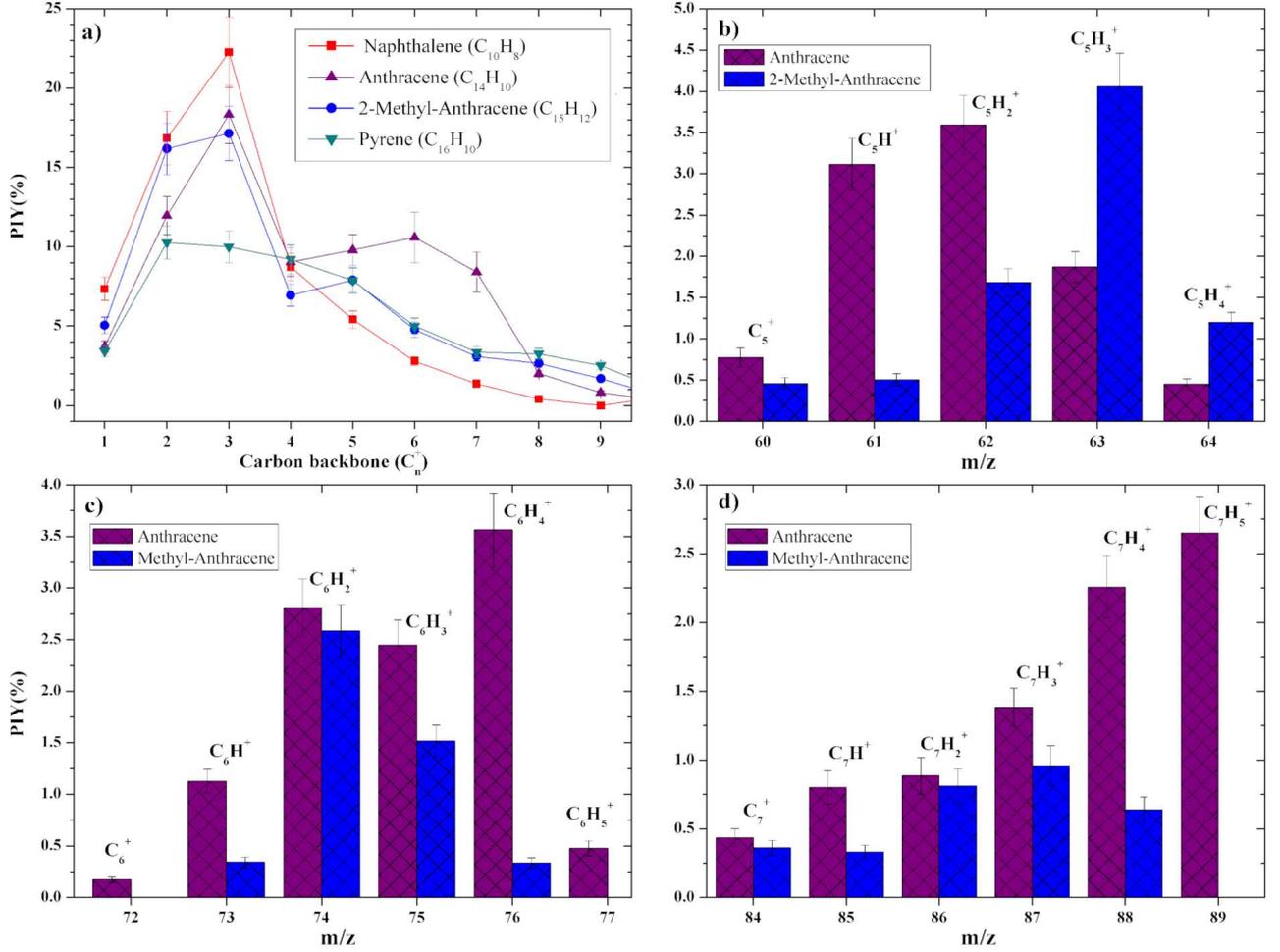}
    \caption{(a) Comparison of the production of  ionic fragments of carbon backbone C$_n^+$ ($n$=1-9)  for each PAH molecule photo-dissociated at 2500 eV. Comparison of the photo-fragmentation of  anthracene and 2-methyl-anthracene at 2500 eV by the ions production in the groups (b) C$_5$H$_{m}^+$ ($m$=0-4), (c) C$_6$H$_{m}^+$ ($m$=0-5) and (d) C$_7$H$_{m}^+$ ($m$=0-6) .}
    \label{fig:5PIYsingchar}
\end{figure*}

Figure \ref{fig:5PIYsingchar} (a) shows the comparison of the singly charged fragments C$_n^+$ ($n$=1-9) produced by X-ray photodissociation of the PAH molecules studied herein. It is possible to see that light C$_n^+$ fragments are more produced by small PAHs, whereas the opposite trend is observed for the heavy ones. For intermediate size fragments ($n$=5-7), the production of ions coming from anthracene is somewhat higher than for the other molecules. In Figure \ref{fig:5PIYsingchar} (b), (c) and (d) we show a comparison between the production of groups of ions C$_5$H$_{m}^+$ ($m$=0-4), (c) C$_6$H$_{m}^+$ ($m$=0-5) and (d) C$_7$H$_{m}^+$ ($m$=0-6) from the  anthracene and 2-methyl-anthracene photo-dissociations, respectively. By analyzing these figures, it is possible to see that the anthracene profile in the C$_5$-C$_7$ range is dominated by a high production of C$_5$H$^+$ (3.1\%), C$_5$H$_2^+$ (3.6\%), C$_6$H$_4^+$ (3.6\%), C$_7$H$_4^+$ (2.3\%) and C$_7$H$_5^+$ (2.7\%). For some of these ions the m/z ratio is the same as for the doubly charged species, that were discriminated in the spectrum by their particular thin width in comparison to the cation peaks. In these cases, similarly to the case of pyrene, the deconvolution signal processing can introduce additional uncertainty to the $PIY$ values. As a consequence, our calculated singly-charged $PIY$s for anthracene in the C$_5$-C$_7$ range could be overestimated, leading to an underestimation of the C$_{10}$-C$_{14}$ dication production for the molecule, with exception of the doubly charged parent ion, which was calculated using the relation with its isotopologue. Nevertheless, the production of C$_{14}$H$_{10}^{2+}$ by photoionization of anthracene triggered by X-rays obtained in this work is significantly higher then the ones obtained by protons (15 keV) and $\alpha$-particle (30 keV) impacts \citep{Postma2010}, as will be discussed in the next section.  

\subsection{Production of Multiply Charged PAHs}
\label{sub:dic}

\begin{figure}
\centering
       \includegraphics[width=8.5cm]{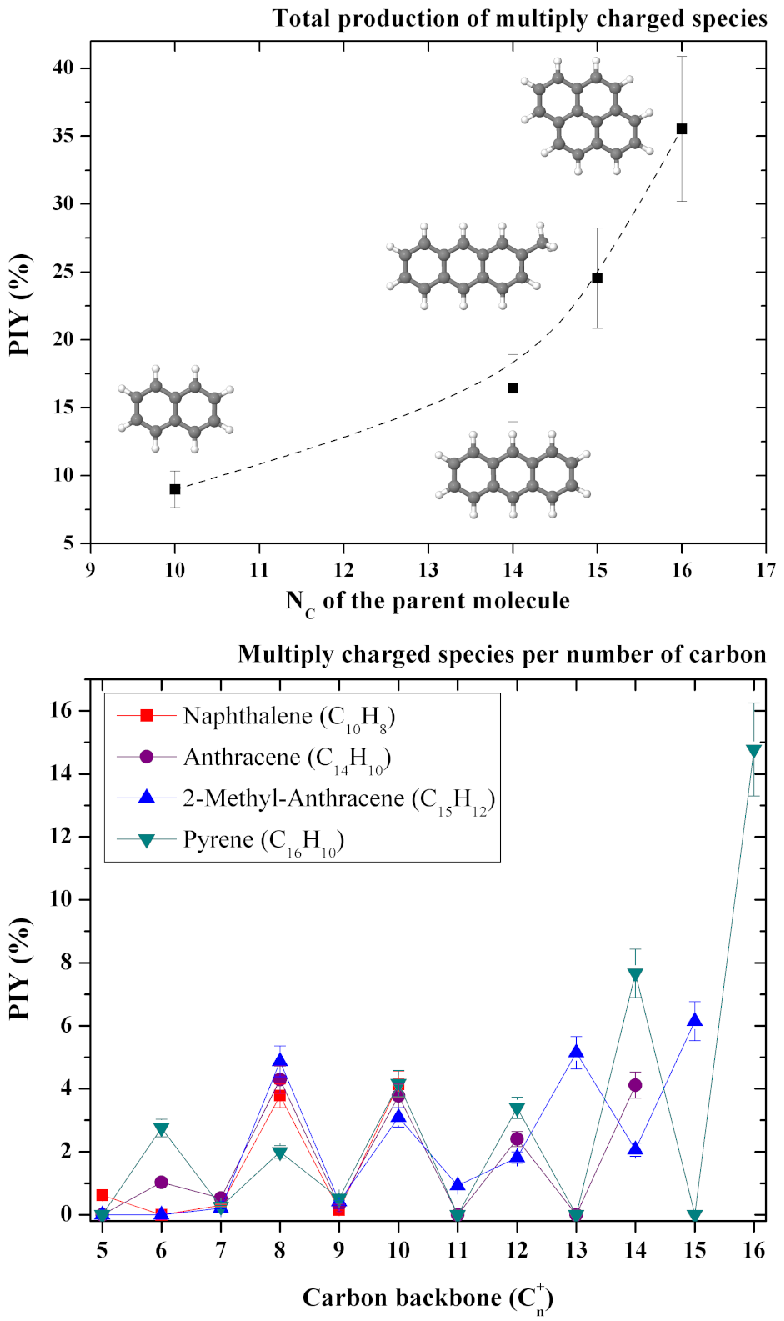}
    \caption{Top: Total $PIY$ (\%) of multiply charged ions produced by the single ionization for each molecule due to photons of 2500 eV. Bottom: Total $PIY$ (\%) of multiply charged ions per size of the carbon backbone, or per groups of fragments containing 5–16 carbon atoms}
    \label{fig:6PIYmultchar}
\end{figure}

The total production of multiply charged species generated by the interaction of a 2500 eV photon with each one of the four PAH-type molecules studied herein is shown in Figure \ref{fig:6PIYmultchar} (top). As the carbon backbone increases, the $PIY$ associated to the formation of multiply charged species also increases. While for naphthalene around 9\% of the photon absorption processes lead to the formation of multiply charged molecular ions, for pyrene this quantity reaches $\sim$36\%, almost four times higher. This result supports the notion that multiple ionization pathways are more prominent for larger PAHs, probably due to a better charge stabilization throughout a larger molecular area. 

In the bottom panel of Figure \ref{fig:6PIYmultchar}, we show the $PIY$ values of multiply charged species as a function of the dication carbon backbone. For the non-methylated PAHs, it is possible to see that the fragments containing an even number of carbon atoms are largely produced in comparison to the odd C$_n$ ones. This result suggests that such electron-poor species are generated mostly from C$_2$H$_2$ loss processes starting from a multiply charged parent ion. For 2-methyl-anthracene, the production of C$_{13}$ multiply charged fragments is high, also indicating that the loss of one C$_2$H$_2$ molecule is a dominant dissociation process. However, the production of C$_{11}$ fragments is considerably negligible, as the ones with smaller odd C$_n$ backbone. This indicates that the production of multiply charged species is more dependent on their intrinsic stabilities, rather than on the characteristics of the particular PAH molecule that is interacting with the 2.5 keV photon. Similar findings have been obtained for the production of C$_n$H$_m^{\pm}$ ions desorbed from condensed hydrocarbon samples after the impact of $^{252}$Cf fission fragments \citep{Fantuzzi2013}. 

\begin{table}
\centering
\caption{Partial Ion Yields ($PIY$) of the singly, doubly and triply charged parent ions of naphthalene (C$_{10}$H$_8$), anthracene (C$_{14}$H$_{10}$), 2-methyl-anthracene (C$_{15}$H$_{12}$), and pyrene (C$_{16}$H$_{10}$) due to photons of 2500 eV. The $PIY$[M]$^{q+}$ (q=2,3) entry stands for the sum of $PIY$[M]$^{2+}$ and $PIY$[M]$^{3+}$, while $PIY$[M]$^{q+}$/$PIY$[M]$^+$ represents the ratio between multiple and single ionization. The $PIY$ values of all species analyzed in this work are shown in Table \ref{tab-all} in the Appendix.}
\label{tab:m}
\begin{tabular}{@{}lcccc@{}} \toprule
\multicolumn{1}{@{}c}{ } & \multicolumn{4}{@{}c}{$PIY$ (\%)}       \\ \cmidrule{2-5}
Label & C$_{10}$H$_8$ & C$_{14}$H$_{10}$ & C$_{15}$H$_{12}$ & C$_{16}$H$_{10}$ \\ \midrule
$PIY$[M]$^+$      & 0.19          & 0.21     & 0.13         & 0.24  \\
$PIY$[M]$^{2+}$     & 0.84        & 1.1       & 1.0       & 2.9   \\
$PIY$[M]$^{3+}$     & 0.47        & 0.29       & --       & 0.65   \\
$PIY$[M]$^{q+}$ (q=2,3) & 1.3    & 1.4       & 1.0       & 3.6   \\ \midrule
\multicolumn{1}{@{}c}{ } & \multicolumn{4}{@{}c}{$PIY$[M]$^{q+}$/$PIY$[M]$^+$}       \\ \cmidrule{2-5}
$q$=2  & 4.4       & 5.2    & 7.7   & 12.1 \\
$q$=3  & 2.5       & 1.4    & --      & 2.7  \\
$q$=2,3  & 6.9     & 6.6    & 7.7    & 14.8  \\
\bottomrule
\end{tabular}
\end{table}

\begin{figure}
\centering
       \includegraphics[width=8.5cm]{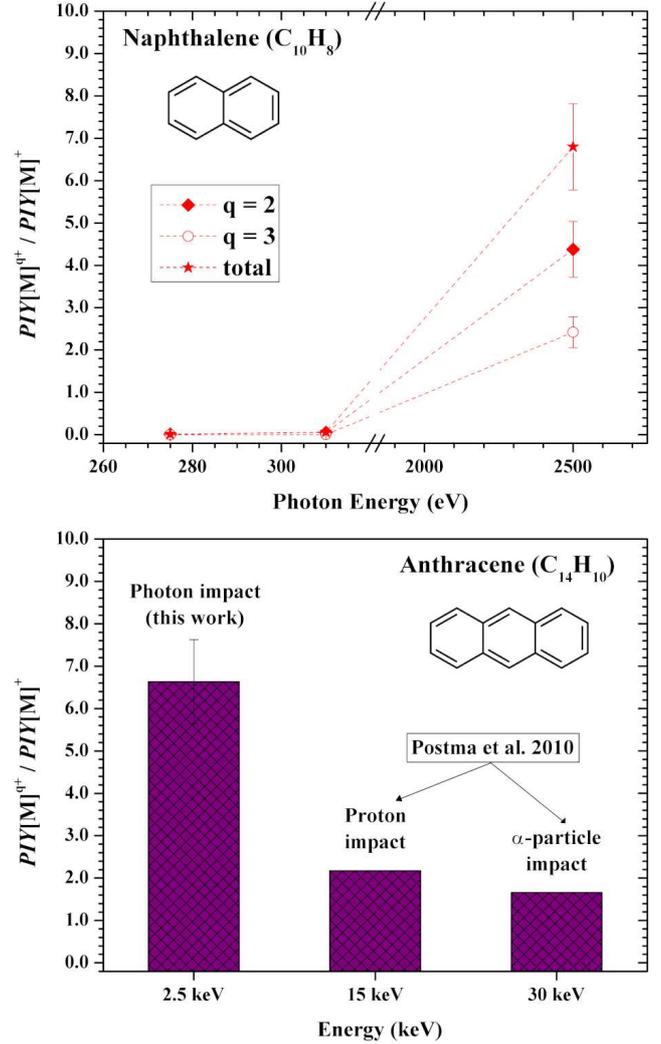}
    \caption{Top: Ratio of the PIYs of multiply charged ions M$^{q+}$ and single charged ions M$^{+}$ for naphthalene as a function of the photon energy. Bottom: Comparison of our result for the anthracene molecule at photon energy of 2500 eV with the \citealt{Postma2010} results using proton and $\alpha$-particle impacts.}
    \label{fig:7}
\end{figure}

The relatively high production of multiply charged PAHs triggered by 2500 eV photons, particularly the yields related to non-dissociative multiple photoionization processes, is of special importance because such values are not observed for energies close to the C1s resonance energy. Two factors influence the tendency observed for higher energies. The first one is related to the $PIY$ of singly charged parent ions, whose values drop down substantially from the C1s resonance energy to energies above. For naphthalene (see Table \ref{tab-all} in the Appendix), the $PIY$ goes from 29.6\% at 275 eV to merely 0.22\% at 2500 eV, which corresponds to a shrinkage of more than 100 times. Similar values are obtained from the other molecules at the latter photon energy, as shown in Table \ref{tab:m}. The second factor is related to an increase of the yield of multiply charged parent ions. For naphthalene (see Table \ref{tab-all} in the Appendix), the sum of C$_{10}$H$_8^{2+}$ and C$_{10}$H$_8^{3+}$ goes from 0.37\% at 275 eV to 1.3\% at 2500 eV, which corresponds to an increase of more than 3.5 times. The preference of multiple over single ionization for high energies becomes even more evident by analyzing the ratio between the $PIY$ values related to both processes. The variation of such ratio, herein labeled as $PIY$[M]$^{q+}$/$PIY$[M]$^+$, with respect to the photon energy for the naphthalene molecule is shown in Figure \ref{fig:7} (top). While $PIY$[M]$^{q+}$/$PIY$[M]$^+$ is practically zero for 275 and 310 eV, this quantity reaches 6.9 ($q$=2,3, see Table \ref{tab:m}) at 2500 eV. For the other molecules (see Table \ref{tab:m}), the ratios are as follows: 6.6 for anthracene, 7.7 for 2-methyl-anthracene, and 14.8 for pyrene. By decomposing $PIY$[M]$^{q+}$/$PIY$[M]$^+$ into contributions arising separately from the doubly and triply charged parent ions, it is possible to see that this ratio is dominated by the M$^{2+}$ yield. These results suggest that the dominance of multiple over single photoionization originated from collisional processes with keV X-rays could lead to an enrichment of multiply charged PAHs in regions where these photons are abundant, such as the circumnuclear envelopes of AGNs. Implications of these results to the chemistry of active galaxies will be discussed in section \ref{sec:ach}. 

Figure \ref{fig:7} (bottom) shows a comparison between the $PIY$[M]$^{q+}$/$PIY$[M]$^+$ ($q$=2,3) ratio obtained in this work for anthracene with photon impact and the ones obtained by \cite{Postma2010} for the same molecule with proton and $\alpha$-particle collisions, with energies of 15 and 30 keV, respectively. Only the beam energies that originated the highest $PIY$[M]$^{q+}$/$PIY$[M]$^+$ ratios among the ones studied by \cite{Postma2010} are presented herein. The collisional processing of PAHs by ions is present in interstellar shocks \citep{Micelotta2010}. For supernovae remnants, such processing also occurs, and it has been studied for large dust grains \citep{Micelotta2018}. Our results suggest that the non-dissociative multiple ionization is significantly more efficient upon the interaction with 2500 eV photons than by collision with protons and $\alpha$-particles, and should take place in regions isolated from the hot gas of these shock waves. Also, this is another indication that such high energy photons could trigger the formation of multiply charged PAHs in the thick tori of AGNs, since they combine a higher selectivity for multiple ionization and a deeper penetrating power than proton and electron interactions \citep{Micelotta2011}.

\subsection{Photoabsorption, Photoionization and Photodissociation Cross Sections of PAHs} \label{sub:cross}

The photoabsorption cross section as a function of the photon energy ($\sigma_{ph-abs} (E)$) of a molecule is an essential parameter to determine the absolute cross sections of photoionization and photodissociation using our $PIY$ results. As mentioned before, we measured the X-ray photodissociation of PAHs at energies of 275, 310 and 2500 eV. 
The values of $\sigma_{ph-abs}$ at energies around the C1s resonance (285 eV) were obtained from the core excitation database of Hitchcock Group (\url{http://unicorn.mcmaster.ca/corex/cedb-title.html}). For the naphthalene molecule, the $\sigma_{ph-abs}$ at 275 eV and 310 eV are 4.4 x 10$^{-2}$ Mb and 10.5 Mb (1Mb = 10$^{-18}$ cm$^{2}$) respectively.

However, to the best of our knowledge, there is neither experimental nor theoretical data on the absolute photoabsorption cross section of PAHs at 2500 eV. Even for benzene, which is the building block of PAHs and one of the most emblematic molecular systems, such measurements have only been made up to 800 eV \citep{Rennie2000}. 

\begin{figure}
\centering
       \includegraphics[width=1.0\columnwidth]{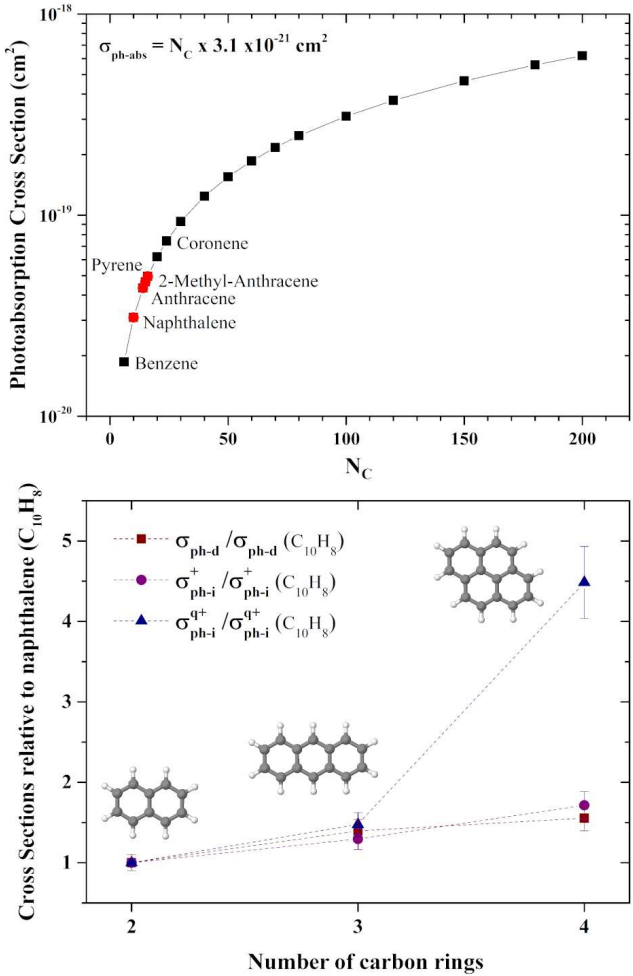}
    \caption{Top: Photoabsorption cross section (cm$^{2}$) at 2500 eV as a function of the number of carbons (N$_{C}$) for each molecule. Bottom: Cross-sections relative to naphthalene as a function of the number of carbon rings.}
    \label{fig:csrel}
\end{figure}

In this work, we obtained the $\sigma_{ph-abs}$ values of the PAH molecules at 2500 eV by multiplying the photoabsorption cross section of the carbon atom (3.1 x 10$^{-21}$ cm$^2$), taken from the literature \citep{Henke1982,Voit1992,Henke1993,Berkowitz2002}, by the number of C atoms (N$_C$) that compose the carbon backbone. The dependence of the absolute photoabsorption cross section with respect to the number of carbon atoms in the backbone of the PAH is shown in Figure \ref{fig:csrel} (top) and the $\sigma_{ph-abs}$ values for the molecules studied herein are presented in Table \ref{tab-sig}. As we can see, the $\sigma_{ph-abs}$ values at 2500 eV do not vary significantly for PAHs with N$_C$= 10 to 16. On the other hand, the photoabsorption cross section at 2500 eV (3.1x10$^{-20}$ cm$
^2$) for the naphthalene molecule is about two orders of magnitude smaller than at 310 eV (1.05x10$^{-17}$ cm$^2$). In fact, it is known that low-energy X-ray photons are more easily absorbed than high-energy X-ray photons \citep{Wilms2000}.  

In the X-ray photon energy range, since the fluorescence yield is usually negligible due to the low carbon atomic number (\citealt{Chen1981}), we may assume that all absorbed photons lead to non-dissociative ionization or dissociative ionization processes. In order words, $\sigma_{ph-abs}$ can be divided into two distinct contributions, namely the non-dissociative photoionization cross section ($\sigma_{ph-i}$) and the photodissociation cross section ($\sigma_{ph-d}$):

\begin{equation} \label{eq:ph-abs}
\sigma_{ph-abs}=  \sigma_{ph-i} + \sigma_{ph-d},
\end{equation}
where $\sigma_{ph-d}$ accounts for all processes in which the ionization induced by the absorption of X-ray photons leads to dissociation.

The non-dissociative photoionization cross section ($\sigma_{ph-i}$) can be partitioned into two terms, depending on the degree of ionization:

\begin{equation} \label{eq:ph-i}
\sigma_{ph-i} = \sigma^+_{ph-i} + \sigma^{q+}_{ph-i} 
\end{equation}
where $\sigma^+_{ph-i}$ is related to the process in which only the singly charged parent ion is produced, while $\sigma^{q+}_{ph-i}$ accounts for non-dissociative multiple ionization processes. The $\sigma_{ph-d}$,  $\sigma^+_{ph-i}$ and $\sigma^{q+}_{ph-i}$ values can be obtained by multiplying $\sigma_{ph-abs}$ and the appropriate $PIY$ quantities, leading to the following equations:   

\begin{subequations}\label{eq:ph-q}
\begin{equation} \label{eq:ph-i1}
\sigma^+_{ph-i} = \sigma_{ph-abs}  \left ( \frac{PIY_{M^+}}{100} \right ), 
\end{equation}
\begin{equation} \label{eq:ph-in}
\sigma^{q+}_{ph-i} = \sigma_{ph-abs} \left (  \sum_{q=2}^{q_{max}}   \frac{PIY_{M^{q+}}}{100}     \right ),
\end{equation}
\begin{equation}
\sigma_{ph-d} = \sigma_{ph-abs} \left (  1 - \frac{PIY_{M^+}}{100} - \sum_{q=2}^{q_{max}}   \frac{PIY_{M^{q+}}}{100}\right),
\end{equation}
\end{subequations}
in which $PIY_{M^+}$ is the partial ion yield value related to the singly charged parent ion, $PIY_{M^{q+}}$ is related to the production of the parent ion of mass $M$ in its charge state $q>1$ and $q_{max}$ is the maximum charge state for which the multiple ionization process is observed. The $PIY$s of singly and multiply charged isotopologues were also taken into account in the calculation of the $\sigma^{+}_{ph-i}$ and $\sigma^{q+}_{ph-i}$ cross sections, respectively.

\begin{table*}
\centering 
\caption{Cross-sections (cm$^{2}$) for photoabsorption  ($\sigma_{ph-abs}$), photodissociation ($\sigma_{ph-d}$), total non-dissociative photoionization ($\sigma_{ph-i}$), single photoionization ($\sigma^{+}_{ph-i}$) and   non-dissociative multiple photoionization ($\sigma^{q+}_{ph-i}$) for the following parent PAH molecules: naphthalene (C$_{10}$H$_{8}$), anthracene (C$_{14}$H$_{10}$), 2-methyl-anthracene (C$_{14}$H$_{9}$CH$_3$) and pyrene (C$_{16}$H$_{10}$), measured at energy of 2500 eV.} 
\label{tab-sig}
\begin{tabular}{l c c c c c}
\\
\toprule
\multirow{2}{*}{Molecule}  & \multicolumn{5}{c}{Cross Section (cm$^{2}$)} \\ 
 & $\sigma_{ph-abs}$	& $\sigma_{ph-d}$ &	$\sigma_{ph-i}$ & $\sigma^+_{ph-i}$ & $\sigma^{q+}_{ph-i}$ 	\\
\midrule
Naphthalene	(C$_{10}$H$_{8}$)  & 3.10$\times 10^{-20}$	& 3.05$\times 10^{-20}$ & 5.35$\times 10^{-22}$ & 6.94$\times 10^{-23}$ & 4.66$\times 10^{-22}$ \\
Anthracene  (C$_{14}$H$_{10}$)  & 4.34$\times 10^{-20}$ & 4.26$\times 10^{-20}$ & 7.78$\times 10^{-22}$ & 8.99$\times 10^{-23}$ & 6.88$\times 10^{-22}$	\\
2-Methyl-Anthracene (C$_{14}$H$_{9}$CH$_3$) &  4.65$\times 10^{-20}$ & 4.59$\times 10^{-20}$ & 6.27$\times 10^{-22}$ & 6.16$\times 10^{-23}$ & 5.66$\times 10^{-22}$	\\
Pyrene (C$_{16}$H$_{10}$) &  4.96$\times 10^{-20}$	& 4.74$\times 10^{-20}$ & 2.20$\times 10^{-21}$ & 1.19$\times 10^{-22}$ & 2.09$\times 10^{-21}$	\\ \bottomrule
\end{tabular}
\end{table*}

The cross sections $\sigma_{ph-abs}$, $\sigma_{ph-d}$, $\sigma_{ph-i}$, $\sigma^+_{ph-i}$ and $\sigma^{q+}_{ph-i}$ values at 2500 eV for the PAHs studied herein are shown in Table \ref{tab-sig}. It is possible to see that the photodestruction of the carbon backbone ($\sigma_{ph-d}$) accounts for the majority of the events that follow photoabsorption, as already anticipated by the $PIY$ analysis. Similarly, the partitioning of $\sigma_{ph-i}$ into $\sigma^+_{ph-i}$ and $\sigma^{q+}_{ph-i}$ contributions also follows the trend already observed in the $PIY$s, which indicates a preference of multiple over single non-dissociative photoionization at 2500 eV. 

The relative weight of $\sigma^{q+}_{ph-i}$ with respect to the increase of the number of rings in the carbon backbone can be traced by analyzing the bottom panel of Figure \ref{fig:csrel}. The plot shows the ratio between $\sigma_{ph-d}$, $\sigma^+_{ph-i}$ and $\sigma^{q+}_{ph-i}$ and their respective values for naphthalene as the number of carbon rings varies from 2 to 4. While the ratios for $\sigma_{ph-d}$ and $\sigma^+_{ph-i}$ exhibit a monotonic growth in this range, $\sigma^{q+}_{ph-i}$ appears to be significantly affected by the increase in the number of carbon rings. This suggests that larger PAHs in X-ray fields can be found in their multiple ionization states. The astrophysical implications of these results to the chemistry of circumnuclear regions of AGNs is discussed in the next section.

\section{Astrophysical Implications: PAHs in AGNs} \label{sec:ach}

\subsection{Photodestruction of PAHs by X-rays in AGNs}

In order to explain the emission of PAHs around AGNs, \cite{Voit1992} proposed that these molecules should be protected by a dense torus surrounding the X-ray source. It is known that the inflowing gas onto a central AGN could form a circumnuclear disk (CND) in which star formation can take place. The CND structure is expected to play an important role in the AGN obscuration, as suggested by starburst disk models \citep{Thompson2005,Ballantyne2008} and high-resolution observations, such as the ones conducted by \cite{Izumi2018}. Moreover, \cite{Kawakatu2008} have shown that the circumnuclear disk might be in coincidence with the putative torus of the standard unification model for AGNs. The authors also predicted that star formation -- and consequently PAH emission -- is more likely to occur in the outer parts of a 100 pc-size torus. Using 8-m class telescopes and Spitzer, \cite{Sales2013} and \cite{Esquej2014} shortened this distance to around a few pc, with resolution of 26 pc. Additionally, using data from ALMA, \cite{Salak2017} mapped a molecular torus by observing the CO(3-2) and CO(1-0) lines, with a radius of $\sim$30 pc, around the central X-ray source. These observations might be in coincidence with the PAH emission from \cite{Sales2013}.

To the best of our knowledge, the photodestruction of PAHs due to interaction with high energy photons ($E > 1000$ eV) has only been studied by extrapolation from indirect measurements. Such estimates were based on PAH experiments with lower energy photons, up to the VUV range \citep{Leach1989, Leach1989b,Jochims1994,Jochims1996,Jochims1999}, and on the photoabsorption profile of the C atom in a photon energy range up to 30 keV \citep{Henke1982,Henke1993}. In addition, the interaction of PAHs and their precursors with photons around the C1s$\rightarrow \pi$* resonance (285 eV) has been explored by several authors \citep{Rennie2000,Boechat2009,Reitsma2014,Reitsma2015,Monfredini2016,Quitian-Lara2018}. However, there is still a lack of data when it comes to investigating the effects of keV photons on these astrophysically relevant molecules.

\begin{table}
\centering
\caption{Some properties of the AGN sources studied herein. $L_X$ stands for the X-ray luminosities integrated from 2-10 keV, while $\xbar{F}_X$ are the respective average X-ray photon fluxes at 2500 eV ($\tau_x = 4.45$) within distances of 20-80 pc from the Seyfert nucleus.} 
\label{tab:agn}
\begin{tabular}{lccc}
\toprule
Source & Type & {$L_X$} & {$\xbar{F}_X$}  \\
                                 &                         & {(eV s$^{-1}$)}                  & {(photons cm$^{-2}$s$^{-1}$)}       \\  \midrule   
Mrk 279$^a$  & Sy1 & 2.48$\times 10^{55}$ & 6.68$\times 10^{8}$  \\
Mrk 334$^b$  & Sy1 & 1.37$\times 10^{54}$                     & 3.68$\times 10^{7}$                   \\
Mrk 3$^b$    & Sy2                     & 1.16$\times 10^{55}$                     & 3.12$\times 10^{8}$                 \\

NGC 5728$^b$ & Sy2                     & 9.42$\times 10^{53}$                    & 2.54$\times 10^{7}$                   \\
NGC 7682$^d$ & Sy2                     & 3.28$\times 10^{53}$                     & 8.82$\times 10^{6}$                \\
NGC 1808$^c$ & Sy2 					   & 3.13$\times 10^{51}$ & 8.41$\times 10^{4}$  \\
\bottomrule
\multicolumn{4}{l}{$^a$ \cite{Vasudevan2009} $^b$ \cite{Shu2007}} \\ 
\multicolumn{4}{l}{$^c$ \cite{Esparza-Arredondo2018} $^d$ \cite{Gu2002}}
\end{tabular}
\end{table}

We applied our experimental data to estimate the stability of PAHs in the circumnuclear regions of six different AGNs (Table \ref{tab:agn}): Mrk 279, Mrk 334, Mrk 3, NGC 5728, NGC 7682, and NGC 1808. These AGNs were taken from the \cite{Sales2010} catalog, and all of them show emission in 8.6 $\mu$m, which is due to ionized PAHs \citep{Draine2001, Draine2007, Sales2010,Sales2013}. The first two entries are Seyfert 1 (Sy1) AGNs, while the others are classified as Seyfert 2 (Sy2). In order to know the X-ray photon flux, $F_X$ (photons cm$^{-2}$ s$^{-1}$), at a given distance from the central source of the AGN and for a given photon energy  $E = h\nu$, we applied, for each object, the following equation:

\begin{equation} \label{eq:flux}
F_{X}= \frac{L_{X}}{4\pi r^{2}h\nu}e^{-\tau_{x}} 
\end{equation}
where $L_x$ is the X-ray luminosity (eV s$^{-1}$) integrated from 2 to 10 keV, $r$  is the distance from the nucleus to a position between 20 to 80 pc inside the dust torus, and $\tau_{x}$ is the X-ray optical depth. Here, the $\tau_{x}$  values were systematically varied from  $\tau_x = 0.0$ to $\tau_x = 10.0$ in order to discuss the importance of shielding to the survival of PAHs in AGNs. These values of $\tau_x$ can be associated with the respective H$_2$ column densities by considering the following equation:

\begin{equation} \label{eq:tau-x}
\tau_x = 2 \sigma_H(E) N_{H_2}
\end{equation}

where $\sigma_H(E)$ is the X-ray photoabsorption cross section per H nucleus given by \cite{Gorti2004}:

\begin{equation} \label{eq:sigmah}
\sigma_H(E) = 1.2 \times 10^{-22} \left ( \frac{E}{1 \, \text{keV}} \right ) ^{-2.594}
\end{equation}

In the context of AGNs, the structure and dynamics of the ionic (H II), atomic (H I) and molecular (H$_2$) gases have been studied by three-dimensional hydrodynamic simulations which include radiative feedback from the central source \citep{Wada2012,Wada2016,Izumi2018}. These studies led to the development of the multi-phase dynamic torus model, in which a combination of dusty outflows, inflows and failed winds give rise to a geometrically thick structure. Dense molecular gases are distributed near the equatorial plane, whereas the atomic gas is more extended along the vertical direction of the disk due to turbulence effects. In our approach, we are considering X-ray obscuration by the mid-plane H$_2$ column density, which would more effectively protect PAHs from photoprocessing.

\begin{figure}
\centering
       \includegraphics[width=1.0\columnwidth]{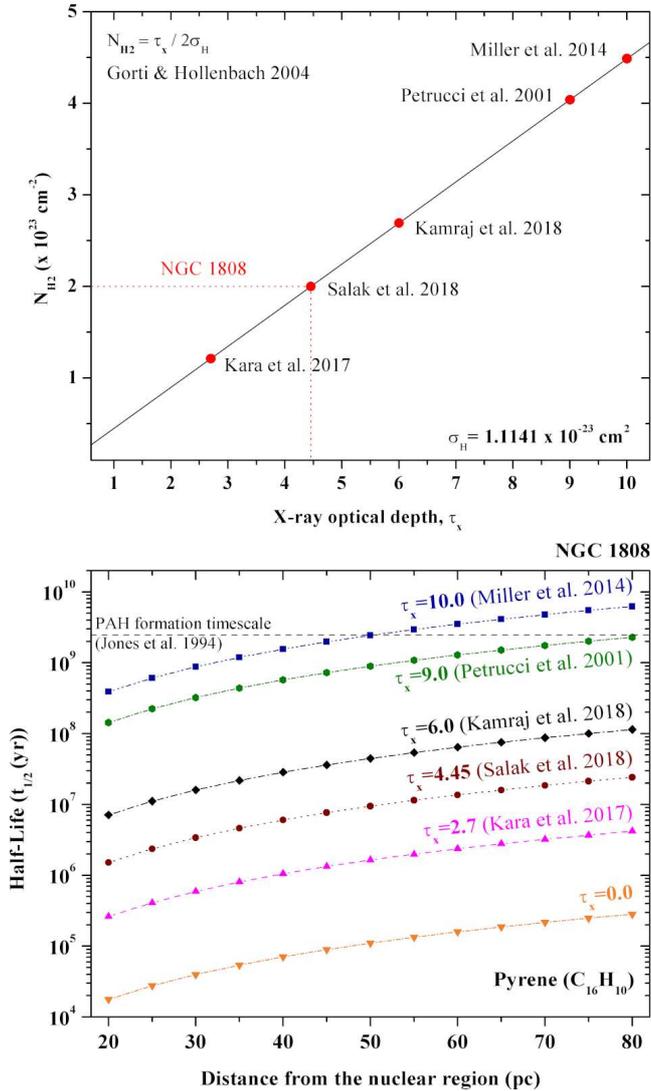}
\caption{Top: Column density of H$_2$ (black line) as a function of the X-ray optical depth, $\tau_x$ ($E = 2.5$ keV, see eq. \ref{eq:sigmah}). The red datapoints are $\tau_x$ values taken from the literature. The datapoint of \citealt{Salak2018} is related to the upper limit of $N_{H_2}$ for NGC 1808. Bottom: Half-life of pyrene (C$_{16}$H$_{10}$) as a function of the distance (pc) from the central source of NGC 1808. Distinct X-ray optical depth values ($\tau_x$) were considered. The horizontal dashed line is the PAH injection timescale estimate of $2.5 \times 10^9$ yr \citep{Jones1994}, which is shown for comparison. See text for details.}
    \label{fig:tau}
\end{figure}

Figure \ref{fig:tau} (top) shows the $N_{H_2}$ values as a function of the X-ray optical depth according to eq. \ref{eq:tau-x} ($E = 2.5$ keV). By taking $N_{H_2} = 2.0 \times 10^{23}$ cm$^{-2}$, which is the upper limit of the H$_2$ column density for NGC 1808 \citep{Salak2018}, we obtain that $\tau_x \sim 4.45$. Since this value is estimated by considering an optically thick molecular mass, it will be used as a reference scenario for the six AGNs studied herein. In addition to the datapoint related to the work of \cite{Salak2018}, we also highlight $\tau_x$ values (and their corresponding $N_{H_2}$ estimates) proposed by different authors for distinct AGN sources. Except for $\tau = 2.7$ \citep{Kara2017}, all other values are higher than the upper limit for $N_{H_2}$ estimated by \cite{Salak2018}.

The photoabsorption (k$_{ph-abs}$), photoionization (k$_{ph-i}$) and photodissociation (k$_{ph-d}$) rates (s$^{-1}$) of each PAH molecule, for a given distance $r$ from the central source, are determined by multiplying the respective cross section, $\sigma_{ph} (E)$, by the photon flux, $F_{x} (E)$: 

\begin{equation} \label{eq:rates}
k_{ph}= \sigma_{ph} (E)F_x(E) 
\end{equation}

where k$_{ph}$ represents the rates (k$_{ph-abs}$, k$_{ph-d}$ or k$_{ph-i}$) and $\sigma_{ph} (E)$ can be the $\sigma_{ph-abs} (E)$, $\sigma_{ph-d} (E)$ or $\sigma_{ph-i} (E)$. By applying eqs. \ref{eq:ph-i1} and \ref{eq:ph-in} to eq. \ref{eq:rates}, it is possible to estimate separately the single ($k^+_{ph-i}$) and multiple ($k^{q+}_{ph-i}$) non-dissociative photoionization rates. We obtained these rates for naphthalene, anthracene, 2-methyl-anthracene and pyrene at $E$ = 2500 eV in the AGN sources studied herein, as shown in Figure \ref{fig:agn-ion}.

From the photodissociation rate, $k_{ph-d}$, it is possible to determine the half-life of a given molecule by considering the exponential decay of an initial molecular abundance $N_{0}$ \citep{Cottin2003}:

\begin{equation}
N(t)=N{_0} \,e ^{ \,-k_{ph-d} \, t}
\end{equation}

The half-life, $t_{1/2}$, is the time required for the abundance to fall by half of its initial value, or $N = N_0/2$. Therefore we can write:

\begin{equation}\label{eq:hl}
t_{1/2}=\frac{ln \, 2}{k_{ph-d}} 
\end{equation}

\begin{figure*}
 \centering
     \includegraphics[width=18cm]{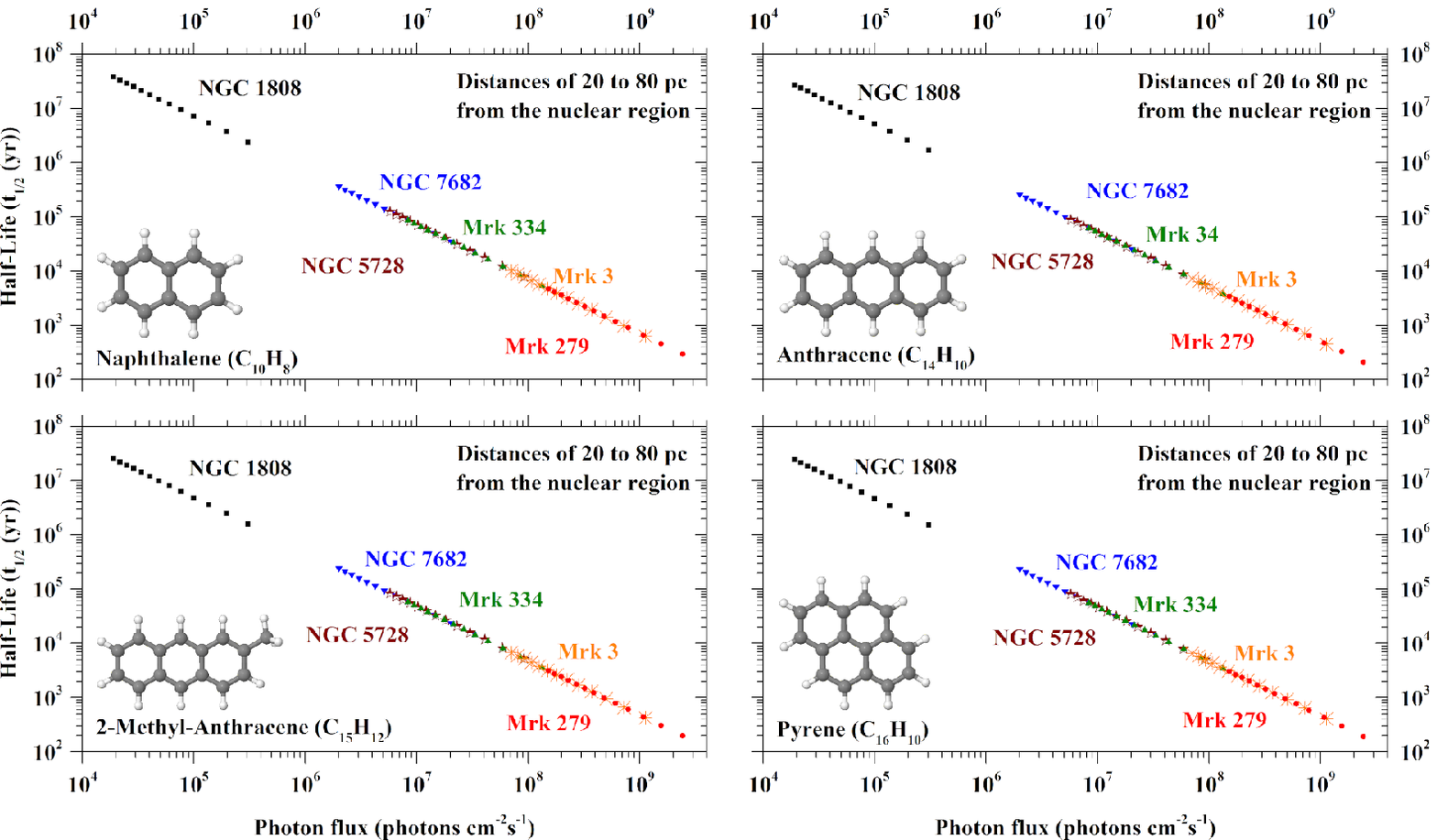}
     \caption{Half-life values of the PAH molecules studied herein after interaction with the X-ray radiation field ($\tau_x = 4.45$) of selected AGNs. Each datapoint refers to a specific distance in the 20-80 pc range from the Seyfert nucleus.}
     \label{fig:agn-hl}
 \end{figure*}
 
Figure \ref{fig:tau} (bottom) shows the half-life estimates of pyrene (C$_{16}$H$_{10}$) for distances between 20 and 80 pc from the nuclear region of NGC 1808. Each curve represents a different value for the X-ray optical depth ($0.0 \leq \tau_x \leq 10.0$). For comparison, injection timescale estimates ($2.5 \times 10^9$ yr, see \citealt{Jones1994}) assuming carbon-rich AGB stars as the primary source of PAHs are also shown. For $\tau_x = 0.0$, the half-lives span from merely $1.8 \times 10^4$ to $ 2.8 \times 10^5$ yr, which are at best four orders of magnitude shorter than the PAH injection estimates. This large discrepancy illustrates the importance of shielding effects by the dusty torus in order to account for the survival of PAHs in such X-ray luminous sources. These values are weakly affected if an optically thin dust model is considered. For $\tau_x = 2.7$  ($N_{H_2} = 1.2 \times 10^{23}$ cm$^{-2}$), the half-life values span from $2.6 \times 10^5 $ to $4.2 \times 10^6$ yr, which are still significantly shorter than the injection times of \cite{Jones1994}. This value of $\tau_x$ was obtained by \cite{Kara2017} after using a thermal Comptonization model to estimate the corona electron temperature of Ark 564, a narrow-line Seyfert 1 AGN.

The half-life estimates are still unsatisfactory even if $\tau_x=$ 4-6 values are considered. For $\tau_x = 4.45$, which is related to the upper limit value of $N_{H_2} = 2 \times 10^{23}$ cm$^{-2}$ for NGC 1808 \citep{Salak2018}, the half-lives are in the range of 1.5$\times 10^{6}$ yr to 2.5$\times 10^{7}$ yr. For $\tau_x = 5.0$ ($N_{H_2} = 2.2 \times 10^{23}$ cm$^{-2}$), the half-lives span from $2.6 \times 10^6 $ to $ 4.2 \times 10^7 $ yr, while for $\tau_x = 6.0 $  ($N_{H_2} = 2.7 \times 10^{23}$ cm$^{-2}$) they range from $ 7.1 \times 10^6 $ to $ 1.1 \times 10^8$ yr. These values of X-ray optical depths are associated with microquasar and AGN coronae. A scattering optical depth value of 5.0, for example, was found for the Comptonizing corona of the microquasar GRS 1915+105 \citep{Ueda2009}, while a $\tau_x = 6.0$ value was used by \cite{Kamraj2018} in conjunction to theoretical constrains from \cite{Petrucci2001} to describe the coronal properties of a sample of \textit{NuSTAR}-observed Seyfert 1 AGNs.

The survival time of pyrene towards the inner region of NGC 1808 is only compared to the PAH injection timescale of \cite{Jones1994} if very high X-ray optical depth values, such as $\tau_x = 9.0$ ($N_{H_2} = 4.0 \times 10^{23}$ cm$^{-2}$) and $\tau_x = 10.0$ ($N_{H_2} = 4.5 \times 10^{23}$ cm$^{-2}$), are considered. For $\tau_x = 9.0$, the half-lives vary from $1.4 \times 10^8 $ to $ 2.3 \times 10^9$ yr, while for $\tau_x = 10.0$ these values span from $3.9 \times 10^8$ to $ 6.2 \times 10^9$ yr. Although these high values of $\tau_x$ were reported in the literature for some objects, such as the AGN corona of NGC 4151 \citep{Petrucci2001} and ultra-luminous X-ray sources (ULXs, see \citealt{Miller2014}), they are overly unrealistic for the purposes of this work. By taking the results of Figure \ref{fig:tau} altogether, the X-ray shielding mechanism by the torus seems to be insufficient to account for the existence of pyrene and other small-size PAHs in the inner regions of AGNs. In fact, given the large differences in the injection and destruction timescales, our results point out that a more sophisticated interplay between PAHs and dust grains should be considered. In the following paragraphs, we briefly discuss two possible, non-exhaustive, scenarios in which dust grains could aid the production and survival of PAHs in the circumnuclear vicinity of AGN sources. Although based on experimental and observational findings described by several authors, the feasibility of these processes is, for practical reasons, still speculative. A quantitative assessment of the role of dust in the survival of PAHs is far beyond the scope of this work.

In the first scenario, namely PAH-to-dust adsorption, we suggest that small-size PAHs formed in the circumnuclear environments of AGNs should be adsorbed onto the dust grains, where they could grow or rebuild themselves by chemical reactions driven by thermal- and/or photo-processing. In the gas-phase, these species could be produced, for instance, following barrierless reactions typical of low-temperature chemistry \citep{Jones2011,Parker2012,Kaiser2015,Lee2019}. On the surface of grains, small-size PAHs could undergo molecular growth by bottom-up chemical reactions, such as the ones described by \cite{Zhao2016,Johansson2018}. Top-down routes of PAH formation, including the path depicted by \cite{Merino2014} involving graphitized grain surfaces exposed to atomic hydrogen,
may also play a role. Large-size PAHs formed either by bottom-up or top-down processes could be desorbed from the grains and ejected into the gas-phase. Although these species are also subject to carbon backbone dissociation due to interaction with the radiation field, a substantial part of their fragments is expected to exhibit vibrational features typical of aromatic hydrocarbons \citep{Micelotta2011}. Similarly, large fragments of PAHs are detected after the interaction of ions with carbonaceous dust analogues, as recently shown by \cite{Pino2019}. As the size of the parent PAH increases, more photon events are necessary in order to dissociate the whole set of fragments featuring the mentioned vibrational modes. As a consequence, the molecular vibrations of such a cascade of fragments would also contribute to the overall IR emission signatures of PAHs detected in AGNs.

Another possible mechanism in which dust grains could contribute to PAH survival, herein mentioned as the PAH-to-dust incorporation mechanism, is the one previously depicted by \cite{Postma2010} to account for the presence of PAHs in supernova ejecta. In this case, small-size PAHs could be incorporated into the growing dust grains, which could function as PAH reservoirs. The detection of PAHs in individual circumsolar graphite grains extracted from meteorites \citep{Bernatowicz1996,Messenger1998} provides experimental evidence for the viability of such a process. However, as mentioned before, our results are not able to estimate the relative contribution of the distinct scenarios mentioned herein to the overall mechanism which allows PAHs to survive in the circumnuclear regions of AGNs, and whose explanation is still unknown. In this perspective, more work is needed to shed a light on this question, as well as to evaluate the importance of the PAH-to-dust adsorption and incorporation processes discussed herein.

In order to compare the half-life results of pyrene in NGC 1808 with the other molecules and AGN sources studied in this work, we obtained $t_{1/2}$ as a function of the photon flux for an X-ray optical depth of $\tau_x = 4.45$. These results are summarized in Figure \ref{fig:agn-hl}. We could not see any significant difference in the half-life values by increasing the size of the carbon backbone. This is probably due to the fact that we spanned only small-size PAH molecules. For medium- and large-size PAHs, although the X-ray photoabsorption cross sections are increased in comparison to the smaller ones (see Figure \ref{fig:csrel}, top panel), a large number of dissociation pathways will have, as products, PAHs with a smaller number of carbon atoms. Ultimately, this will contribute to increase the half-life of PAHs in our AGN sources. A similar consideration was done by \cite{Micelotta2011} when assuming that the PAH is destroyed by cosmic rays only if the dissociation is followed by the ejection of at least 1/3 of the initial PAH carbon content. 

From Figure \ref{fig:agn-hl}, it is also possible to see that the half-life values are severely affected by the X-ray photon flux experienced by the molecules at a given distance from the central region of the AGN. In spite of considering attenuation of the X-ray radiation field by a dusty torus with moderate optical depth, the lifetime of PAHs spanned values from 10$^8$ to 10$^2$ yr. Even in the best-case scenario, the half-life is still shorter than injection timescale estimates of \cite{Jones1994} by a factor of $\sim$20. 

The highest $t_{1/2}$ values ($10^7$-$10^8$ yr) are obtained for NGC 1808, which also presents the smallest X-ray luminosity (3.13 $\times$ 10$^{51}$ eV s$^{-1}$, \citealt{Esparza-Arredondo2018}) among the AGNs studied herein and, consequently, the smallest average X-ray photon flux at 2500 eV (8.41$\times 10^{4}$  photons cm$^{-2}$ s$^{-1}$). This nearby barred starburst, which is located at a distance of 10.8 Mpc \citep{Tully1988}, is known to have molecular gas outflow from a compact (r < 200 pc) circumnuclear disk, as well as a 500 pc gaseous ring \citep{Salak2016a}. A dynamically driven evolutionary scenario, in which star formation is triggered by the gravitational collapse and cloud-cloud collisions that follow molecular cloud accretion onto this 500 pc ring, was proposed by \cite{Salak2017}. More recently, the authors have also identified new dense gas tracer lines, which corroborates the existence of a velocity gradient in the outflow direction \citep{Salak2018}. PAH emission lines were observed in this object even at small distances (26 pc) from the Seyfert nucleus \citep{Sales2013}, for which our half-life values are merely 10$^6$ yr. 

In contrast to NGC 1808, the shortest half-life times ($\sim$10$^2$ yr) are obtained for Mrk 279, which presents the highest average X-ray photon flux at 2500 eV (6.68$\times 10^{8}$ photons cm$^{-2}$ s$^{-1}$). The absorption features of this very luminous Sy1 galaxy, which is located at 128.6 Mpc \citep{Scott2004}, are consistent with the presence of a warm absorber, most likely a dusty torus \citep{Ebrero2010}. Moreover, weak PAH emission features have been independently observed in Mrk 279 by different groups \citep{Santos-Lleo2001,Sales2010}.  

The determined half-lives of PAHs in the remaining AGN sources span from $\sim$10$^3$ yr to $\sim$10$^6$ yr. Once more, these values suggest that a more complex interplay between PAHs and dust grains should be present in order to account for the survival and detection of PAHs in AGNs.

\subsection{Multiple Ionization of PAHs in AGNs}

\begin{figure*}
\centering
       \includegraphics[width=\textwidth]{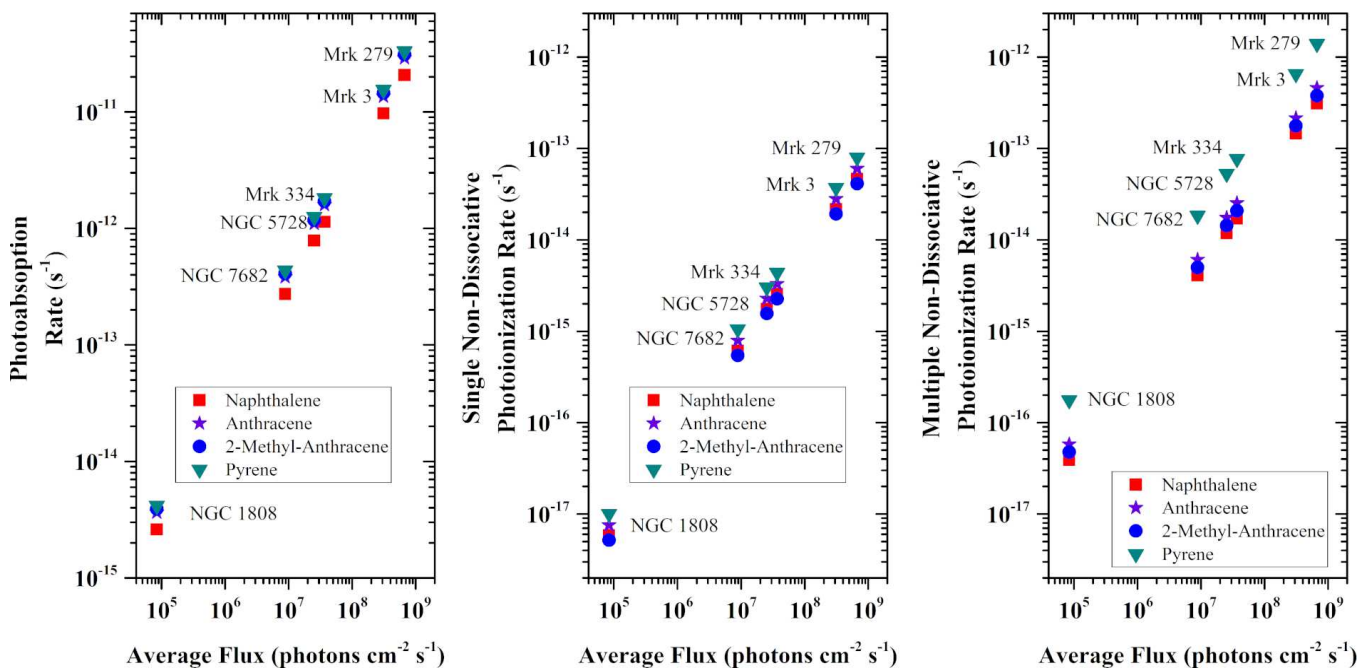}
    \caption{Photoabsorption and photoionization rates (s$^{-1}$) of the PAHs studied herein as a function of the average photon flux $\xbar{F}_X$ of the distinct AGN sources. The $\xbar{F}_X$ values are taken at 2500 eV ($\tau_x = 4.45$) within distances of 20-80 pc from the Seyfert nucleus.}
    \label{fig:agn-ion}
\end{figure*}

The strong X-ray radiation fields that are generated in the central regions of AGNs can easily destroy PAHs, as discussed in the previous section. However, PAHs are substantially formed in such environments, and a conclusive solution for this apparent dichotomy is, to the best of our knowledge, still unknown. In fact, observation of PAH emission in the kpc-scaled circumnuclear vicinity of AGNs have been commonly used to track the presence of star forming activity \citep{Peeters2004,Stierwalt2014,Alonso-Herrero2014,Esparza-Arredondo2018}. This is due to the fact that the UV radiation emitted from young massive O- and B-stars is able to vibrationally excite PAHs \citep{Peeters2004}. The underlying mechanism was recently studied by \cite{Marciniak2015}, which have shown that UV-excited PAHs undergo ultrafast non-adiabatic relaxation through internal conversion processes that couple the electronic and vibrational degrees of freedom. The mid-IR emissions coming from excited PAHs are used as star formation tracers in AGNs because emission from the accretion process contaminates the traditional tracers, such as UV emission, H $\alpha$ and Pa $\alpha$. 

In addition to the destruction pathway, \cite{Jensen2017} have recently shown that the AGN Seyfert nucleus could act as a central excitation source of PAHs within 10-500 pc from the nuclear region. According to their findings, the interaction of shielded PAHs with the radiation field emitted from the AGN can vibrationally excite such molecules, which could undermine the PAH emission as a star forming tracer within the kpc distance range from the central region. 

Besides destruction and vibrational excitation, our results point to a third fate of the PAH molecules that are formed within the inner regions of AGNs. As we show in the previous sections, X-ray photons in the keV range seem to produce multiply ionized PAHs more efficiently than processes associated with proton and $\alpha$-particle impact. Since the removal of two electrons from the aromatic moiety weakens both the C$-$H bonding and the cohesion of the carbon skeleton, the production of such metastable species could enhance fragmentation by activating Coulomb explosion dissociation pathways \citep{Voit1992}. This is corroborated by our results with 2500 eV photons, which show a plethora of low-mass fragments in significantly yield when compared to the mass spectra taken at lower photon energies (see Table \ref{tab-all} for comparison between the mass spectra of naphthalene at 275, 310 and 2500 eV). On the other hand, our results also suggest that the production of stable multiply charged ions increases as a function of N${_C}$. In contrast to doubly charged diatomics, for which metastability is achieved through the formation of strong covalent bonding aided by polarization effects \citep{Fantuzzi2017}, the charge density in polyatomic species can be distributed through a large number of atoms \citep{Cohen1988}. Therefore, by increasing the carbon backbone of the PAH, it is expected that the respective dication (or even higher order charge states) is more easily stabilized through charge alternation. 

Finally, we discuss the possibility of forming multiply charged PAHs in the surroundings of AGNs. Figure \ref{fig:agn-ion} shows a comparison between the photoabsorption rate ($k_{ph-abs}$), the single non-dissociative photoionization rate ($k_{ph-i}^+$) and the multiple non-dissociative photoionization rate ($k_{ph-i}^{q+}$) of the PAHs studied herein in the selected AGN sources. In order to show only one datapoint for each AGN source, we replaced the X-ray photon flux ($F_X$) of eq. \ref{eq:rates}, which depends on the distance from the central Seyfert nucleus, by the average X-ray photon flux ($\xbar{F}_X$) shown in Table \ref{tab:agn}, which is the average flux within 20-80 pc for each one of the sources. It is possible to see that, irrespective of the source, the $k_{ph-i}^{q+}$ values are significantly greater than the corresponding $k_{ph-i}^{+}$ ones, evidencing the high tendency of these systems to form multiply charged states. In addition, it is possible to see that the multiple non-dissociative photoionization rate of pyrene is remarkably higher than the ones of the other PAHs. These results suggest that stable multiply charged PAHs could be formed in the circumnuclear regions of AGNs.

Neutral and ionized PAHs present large differences in the relative intensities of their resulting IR spectra \citep{Allamandola1999,Tielens2008}. However, there is still no evidence for a unambiguous identification between singly and doubly charged PAHs in the ISM \citep{Zhen2017}. The precise determination of the charge state of PAHs is a relevant astrophysical feature, which is also corroborated by our results. Further studies aiming at the spectral signature of these multiply charged species should be developed in the nearby future.

\section{Conclusions}
\label{sec:summary}

In this work, we examined the photoionization and photodissociation profiles of selected polycyclic aromatic hydrocarbons (PAHs) upon their interaction with soft and tender X-rays. For this purpose, mass spectra of naphthalene (C$_{10}$H$_{8}$), anthracene (C$_{14}$H$_{10}$), 2-methyl-anthracene (C$_{14}$H$_{9}$CH$_{3}$, or C$_{15}$H$_{12}$) and pyrene (C$_{16}$H$_{10}$) were obtained for energies below (275 eV) and above (310 eV) the C1s resonance features of these PAHs, and at 2500 eV. The results are discussed in the context of the chemistry of the circumnuclear regions of AGNs, which is the focus of our study. The measurements were performed at the Brazilian Synchrotron Light Laboratory (LNLS) using time-of-flight mass-spectrometry and photoelectron-photoion coincidence techniques.

Two processes emerge as the main channels after excitation and photoionization of the molecules: the breakage of the carbon backbone, in which singly charged ions with m/z lower than half of the parent molecule are mostly produced, and the formation of multiply charged ions, for which the initial PAH carbon content is predominantly preserved.  Multiple non-dissociative photoionization is more pronouncedly activated if the photon energy is increased to the keV region. The same trend is observed if the carbon backbone size of the molecule is expanded. In fact, multiple ionization seems to be significantly more efficient upon the interaction with 2500 eV photons than by collision with high energy protons and $\alpha$-particles. On the other hand, methylation of the PAH does not seem to particularly affect the formation of multiply charged species in the keV region in comparison to its non-methylated analog. The distribution of multiply charged species that are produced upon interaction with a 2.5 keV photon seems to be more dependent on their intrinsic stabilities, rather than on the characteristics of the PAH parent molecule. Taken together, these results could indicate that high energy photons are able to trigger the formation of multiply charged PAHs in the surroundings of AGNs.

By taking our $PIY$ results, we could determine the photoionization and photodissociation cross sections of the PAH molecules at 2500 eV. These values were used to estimate the photoionization and photodissociation rates of PAHs in the circumnuclear regions (20-80 pc) of six AGN sources with distinct X-ray fluxes. From the photodissociation rates, we could estimate the half-lives of those molecules for different optical depth values of the X-ray photon flux. These values were compared to the PAH injection timescale ($2.5 \times 10^9$ yr) described by \cite{Jones1994} assuming that the main sources of PAHs are carbon-rich AGB stars. In spite of considering attenuation of the X-ray radiation field by a dusty torus associated with an H$_2$ column density of $2 \times 10^{23}$ cm$^{-2}$ ($\tau = 4.45$), the lifetime of PAHs spanned values from 10$^8$ yr to merely 10$^2$ yr. These results may indicate that, in order to circumvent molecular destruction, a more sophisticated interplay between PAHs and dust grains should be considered. In this perspective, we briefly describe two possible scenarios in which grains could assist in the survival of PAHs.

We could not see any significant difference in the half-life values by increasing the size of the carbon backbone. This is probably due to the fact that we spanned only small-size PAH molecules ($10 \leq \text{N}_C \leq 16$). In addition, we show that the multiple photoionization rates are significantly greater than the single ones, irrespective of the AGN source. These results suggest that an enrichment of multiply charged ions caused by X-ray photoselection can occur in AGNs. The precise determination of the charge state of PAHs based on specific spectral signatures should be developed in order to confirm this photoselectivity mechanism.

\section*{Acknowledgements}

We would like to thank the staff of the Brazilian Synchrotron Light Laboratory (LNLS)/CNPEM and the financial support from the Conselho Nacional de Desenvolvimento Científico e Tecnológico (CNPq). This study was financed in part by the Coordenação de Aperfeiçoamento Pessoal de Nível Superior (CAPES) - Finance Code 001. We thank the anonymous reviewer for the valuable comments. F.F. acknowledges the hospitality of the Pacific Northwest National Laboratory (PNNL) and the financial support of the Alexander von Humboldt (AvH) Foundation within the program Capes-Humboldt Research Fellowship for postdoctoral researchers, during the period when this manuscript was being prepared for publication. 



\bibliographystyle{mnras}
\bibliography{references}

\appendix

\section{Partial Ion Yields}

\begin{table*}
\centering
\caption{\textit{PIY (\%)} of each molecule, organized per number of carbon backbone, photon energy and charge state.}
\label{tab-all}
\begin{tabular}{lccccccccc}
\toprule
\multirow{3}{*}{m/z} & \multirow{3}{*}{C$_n$} & \multirow{3}{*}{Charge} & \multirow{3}{*}{Attribution} & \multicolumn{3}{c}{Naphthalene} & Anthracene  & 2-Methyl-Anthracene  & Pyrene  \\ 
 &  &  &  & \multicolumn{3}{c}{(C$_{10}$H$_8$)} & (C$_{14}$H$_{10}$) & (C$_{15}$H$_{12}$) & (C$_{16}$H$_{10}$) \\ 
 &  &  &  & 275 eV & 310 eV & 2500 eV & 2500 eV & 2500 eV & 2500 eV \\ \midrule
1	&	0	&	1	&	 H$^+$            	&	4.1	&	16.0	&	24.3	&	7.4	&	6.2	&	 7.3 \\
2	&	0	&	1	&	 H$_2^+$           	&	   --   	&	0.33	&	0.95	&	0.51	&	0.77	&	 0.46 \\
12	&	1	&	1	&	 C$^+$            	&	0.29	&	1.2	&	2.8	&	1.4	&	0.83	&	 0.86 \\
13	&	1	&	1	&	 CH$^+$           	&	0.20	&	0.85	&	1.7	&	0.81	&	0.81	&	 1.1 \\
14	&	1	&	1	&	 CH$_2^+$          	&	   --   	&	0.32	&	0.74	&	0.42	&	0.87	&	 0.65 \\
15	&	1	&	1	&	 CH$_3^+$          	&	0.36	&	1.3	&	1.6	&	1.1	&	2.6	&	 0.84 \\
16	&	1	&	1	&	 CH$_4^+$          	&	   --   	&	   --   	&	0.51	&	  --   	&	  --   	&	     \\
24	&	2	&	1	&	 C$_2^+$           	&	   --   	&	0.56	&	1.9	&	1.0	&	0.72	&	 1.1 \\
25	&	2	&	1	&	 C$_2$H$^+$          	&	0.42	&	1.5	&	2.7	&	1.9	&	1.1	&	 1.1 \\
26	&	2	&	1	&	 C$_2$H$_2^+$         	&	2.8	&	8.0	&	8.1	&	5.3	&	4.6	&	 5.6 \\
27	&	2	&	1	&	 C$_2$H$_3^+$         	&	2.6	&	4.2	&	4.2	&	3.8	&	9.8	&	 2.5 \\
36	&	3	&	1	&	 C$_3^+$           	&	0.45	&	2.1	&	2.2	&	1.5	&	0.60	&	 1.3 \\
37	&	3	&	1	&	 C$_3$H$^+$          	&	2.6	&	5.7	&	6.4	&	6.4	&	5.0	&	 3.1 \\
38	&	3	&	1	&	 C$_3$H$_2^+$         	&	3.1	&	4.9	&	10.2	&	4.4	&	2.3	&	 2.2 \\
39	&	3	&	1	&	 C$_3$H$_3^+$         	&	3.4	&	3.3	&	3.4	&	6.0	&	9.2	&	 3.4 \\
40	&	3	&	1	&	 C$_3$H$_4^+$         	&	0.94	&	   --   	&	   --   	&	  --   	&	  --   	&	  --   \\
48	&	4	&	1	&	 C$_4^+$           	&	0.40	&	0.59	&	0.70	&	0.56	&	  --   	&	 0.47 \\
49	&	4	&	1	&	 C$_4$H$^+$          	&	0.91	&	2.4	&	1.7	&	1.8	&	0.97	&	 2.3 \\
50	&	4	&	1	&	 C$_4$H$_2^+$         	&	5.6	&	7.5	&	4.8	&	4.8	&	2.7	&	 4.9 \\
51	&	4	&	1	&	 C$_4$H$_3^+$         	&	4.8	&	4.7	&	0.94	&	1.8	&	2.7	&	 1.6 \\
52	&	4	&	1	&	 C$_4$H$_4^+$         	&	1.7	&	    --  	&	0.5	&	 --     	&	0.53	&	  --   \\
53	&	4	&	1	&	 C$_4$H$_5^+$         	&	0.28	&	   --   	&	   --   	&	  --   	&	  --   	&	  --   \\
60	&	5	&	1	&	 C$_5^+$           	&	0.91	&	1.3	&	0.55	&	0.77	&	0.46	&	 1.3 \\
61	&	5	&	1	&	 C$_5$H$^+$          	&	1.2	&	3.5	&	1.4	&	3.1	&	0.50	&	 2.4 \\
62	&	5	&	1	&	 C$_5$H$_2^+$         	&	3.4	&	3.5	&	1.1	&	3.6	&	1.7	&	 2.6 \\
63	&	5	&	1	&	 C$_5$H$_3^+$         	&	4.2	&	2.7	&	1.7	&	1.9	&	4.1	&	 1.4 \\
64	&	5	&	1	&	 C$_5$H$_4^+$         	&	3.8	&	0.55	&	0.66	&	0.45	&	1.2	&	 0.21 \\
72	&	6	&	1	&	 C$_6^+$           	&	   --   	&	   --   	&	  --    	&	0.17	&	   --  	&	 0.14 \\
73	&	6	&	1	&	 C$_6$H$^+$          	&	 --  	&	 --  	&	 --  	&	1.1	&	0.34	&	 1.0 \\
74	&	6	&	1	&	 C$_6$H$_2^+$         	&	0.68	&	1.5	&	0.80	&	2.8	&	2.6	&	 1.8 \\
75	&	6	&	1	&	 C$_6$H$_3^+$         	&	2.4	&	2.5	&	1.2	&	2.4	&	1.5	&	 1.7 \\
76	&	6	&	1	&	 C$_6$H$_4^+$         	&	2.3	&	1.8	&	0.80	&	3.6	&	0.34	&	 0.40 \\
77	&	6	&	1	&	 C$_6$H$_5^+$         	&	1.3	&	0.76	&	  --    	&	0.48	&	  --   	&	  --   \\
78	&	6	&	1	&	 C$_6$H$_6^+$         	&	1.2	&	0.40	&	  --    	&	 --     	&	  --   	&	  --   \\ 
84	&	7	&	1	&	 C$_7^+$           	&	   --   	&	0.37	&	0.22	&	0.44	&	0.36	&	 0.67 \\
85	&	7	&	1	&	 C$_7$H$^+$          	&	0.44	&	0.52	&	0.35	&	0.80	&	0.25	&	 1.0 \\
86	&	7	&	1	&	 C$_7$H$_2^+$         	&	0.70	&	1.2	&	0.40	&	0.91	&	0.81	&	 1.7 \\
87	&	7	&	1	&	 C$_7$H$_3^+$         	&	1.0	&	0.61	&	0.41	&	1.4	&	0.96	&	  --   \\
88	&	7	&	1	&	 C$_7$H$_4^+$         	&	0.37	&	0.36	&	 --     	&	2.3	&	0.64	&	  --   \\
89	&	7	&	1	&	 C$_7$H$_5^+$         	&	0.50	&	0.36	&	 --     	&	2.7	&	--	&	 --    \\
97	&	8	&	1	&	 C$_8$H$^+$          	&	    --  	&	0.14	&	  --    	&	0.68	&	0.58	&	 0.36 \\
98	&	8	&	1	&	 C$_8$H$_2^+$         	&	0.48	&	0.59	&	0.41	&	0.57	&	0.72	&	 0.76 \\
99	&	8	&	1	&	 C$_8$H$_3^+$         	&	0.45	&	0.34	&	   --   	&	0.76	&	0.53	&	 --    \\
100	&	8	&	1	&	 C$_8$H$_4^+$         	&	0.32	&	0.18	&	   --   	&	 --    	&	0.37	&	  --   \\
101	&	8	&	1	&	 C$_8$H$_5^+$         	&	0.97	&	0.36	&	  --    	&	 --     	&	  --   	&	 2.2 \\
102	&	8	&	1	&	 C$_8$H$_6^+$         	&	1.8	&	0.55	&	  --    	&	 --     	&	   --  	&	  --   \\
103	&	8	&	1	&	 C$_8$H$_7^{+}$        	&	0.78	&	  --    	&	  --    	&	   --  	&	 --  	&	   --  \\
\bottomrule
\end{tabular}
\end{table*}

\begin{table*}
\centering
\label{my-label}
\begin{tabular}{@{}lccccccccc@{}}
\multicolumn{10}{c}{\textbf{Table \textbf{A1}}. Continued.} \\
\toprule
\multirow{2}{*}{m/z} & \multirow{2}{*}{C$_n$} & \multirow{2}{*}{Charge} & \multirow{2}{*}{Attribution} & \multicolumn{3}{c}{Naphthalene} & Anthracene  & 2-Methyl-Anthracene  & Pyrene  \\ 
  &  &  &  & 275 eV & 310 eV & 2500 eV & 2500 eV & 2500 eV & 2500 eV \\ \midrule
108	&	9	&	1	&	 C$_9^+$           	&	  --    	&	  --    	&	  --    	&	0.34	&	0.14	&	 0.35 \\
109	&	9	&	1	&	 C$_9$H$^+$          	&	   --   	&	  --    	&	  --    	&	0.47	&	0.41	&	 0.70 \\
110	&	9	&	1	&	 C$_9$H$_2^+$         	&	  --    	&	  --    	&	 --     	&	--	&	0.46	&	 1.1 \\
111	&	9	&	1	&	 C$_9$H$_3^+$         	&	   --   	&	  --    	&	 --     	&	  --   	&	0.18	&	 0.40 \\
113	&	9	&	1	&	 C$_9$H$_4^+$         	&	    --  	&	   --   	&	  --    	&	  --   	&	0.16	&	 --    \\
114	&	9	&	1	&	 C$_9$H$_5^+$         	&	   --   	&	  --    	&	  --    	&	  --   	&	0.15	&	   --  \\
115	&	9	&	1	&	 C$_9$H$_6^+$         	&	  --    	&	  --    	&	 --     	&	  --   	&	0.12	&	  --  \\
116	&	9	&	1	&	 C$_9$H$_7^+$         	&	 --     	&	 --     	&	 --      	&	  --   	&	0.09	&	   --  \\
121	&	10	&	1	&	 C$_{10}$H$^+$         	&	   --   	&	  --    	&	  --    	&	  --   	&	 --    	&	 0.18 \\
122	&	10	&	1	&	 C$_{10}$H$_2^+$        	&	  --    	&	 --     	&	 --     	&	0.32	&	0.25	&	 0.50 \\
123	&	10	&	1	&	 C$_{10}$H$_3^+$        	&	  --    	&	 --    	&	  --    	&	   --  	&	  --   	&	 0.29 \\
126	&	10	&	1	&	 C$_{10}$H$_6^+$        	&	3.1	&	0.65	&	   --   	&	 --     	&	0.33	&	  --   \\
127	&	10	&	1	&	 C$_{10}$H$_7^+$        	&	2.0	&	1.7	&	0.34	&	 --    	&	   --  	&	  --   \\
128	&	10	&	1	&	 C$_{10}$H$_8^+$        	&	26.4	&	5.4	&	0.19	&	 --    	&	  --   	&	  --   \\
129	&	10	&	1	&	 $^{13}$CC$_9$H$_8^+$      	&	3.0	&	0.53	&	0.03	&	  --   	&	  --   	&	  --   \\
138	&	11	&	1	&	 C$_{11}$H$_6^+$        	&	  --    	&	  --    	&	  --    	&	  --   	&	0.12	&	   --  \\
139	&	11	&	1	&	 C$_{11}$H$_7^+$        	&	  --    	&	  --    	&	  --    	&	0.24	&	0.17	&	   --  \\
140	&	11	&	1	&	 C$_{11}$H$_8^+$        	&	  --    	&	  --    	&	  --    	&	   --  	&	0.12	&	   --  \\
150	&	12	&	1	&	 C$_{12}$H$_6^+$        	&	   --   	&	  --    	&	  --    	&	  --   	&	0.04	&	  --   \\
151	&	12	&	1	&	 C$_{12}$H$_7^+$        	&	  --    	&	  --    	&	  --    	&	   --  	&	0.16	&	   --  \\
152	&	12	&	1	&	 C$_{12}$H$_8^+$        	&	  --   	&	  --    	&	  --    	&	   --  	&	0.12	&	  --   \\
153	&	12	&	1	&	 C$_{12}$H$_9^+$        	&	  --    	&	  --    	&	  --    	&	 --    	&	0.06	&	   --  \\
163	&	13	&	1	&	 C$_{13}$H$_7^+$        	&	  --    	&	   --   	&	  --    	&	 --    	&	0.19	&	  --   \\
164	&	13	&	1	&	 C$_{13}$H$_8^+$        	&	  --    	&	  --    	&	  --    	&	 --    	&	0.38	&	   --  \\
165	&	13	&	1	&	 C$_{13}$H$_9^+$        	&	  --    	&	  --    	&	 --     	&	   --  	&	0.35	&	  --   \\
166	&	13	&	1	&	 C$_{13}$H$_{10}^+$       	&	  --    	&	 --     	&	 --     	&	  --   	&	0.12	&	   --  \\
177	&	14	&	1	&	 C$_{14}$H$_9^+$        	&	  --    	&	  --    	&	 --     	&	0.20	&	   --  	&	  --   \\
178	&	14	&	1	&	 C$_{14}$H$_{10}^+$       	&	  --    	&	  --    	&	 --     	&	0.21	&	  --   	&	  --   \\
190	&	15	&	1	&	 C$_{15}$H$_{10}^+$       	&	  --    	&	   --   	&	   --   	&	  --   	&	0.15	&	   --  \\
191	&	15	&	1	&	 C$_{15}$H$_{11}^+$       	&	 --     	&	   --   	&	  --   	&	  --   	&	0.17	&	   --  \\
192	&	15	&	1	&	 C$_{15}$H$_{12}^+$       	&	  --    	&	  --    	&	 --     	&	 --    	&	0.13	&	  --   \\
200	&	16	&	1	&	 C$_{16}$H$_8^+$        	&	  --    	&	 --     	&	 --     	&	  --   	&	  --   	&	 0.25 \\
201	&	16	&	1	&	 C$_{16}$H$_9^+$        	&	  --    	&	 --     	&	 --     	&	  --   	&	  --   	&	 0.24 \\
202	&	16	&	1	&	 C$_{16}$H$_{10}^+$       	&	  --    	&	  --    	&	 --     	&	  --   	&	  --   	&	 0.24 \\
31	&	5	&	2	&	 C$_5$H$_2^{2+}$        	&	0.01	&	0.65	&	0.62	&	 --     	&	  --   	&	  --   \\
36.5	&	6	&	2	&	 C$_6$H$^{2+}$         	&	  --    	&	  --    	&	 --     	&	  --   	&	  --   	&	 0.74 \\
37.5	&	6	&	2	&	 C$_6$H$_3^{2+}$        	&	   --   	&	0.13	&	  --    	&	   --  	&	  --   	&	 0.89 \\
38.5	&	6	&	2	&	 C$_6$H$_5^{2+}$        	&	0.07	&	0.16	&	  --    	&	1.0	&	  --   	&	 1.1 \\
39.5	&	6	&	2	&	 C$_6$H$_7^{2+}$        	&	   --   	&	0.17	&	 --     	&	  --   	&	  --   	&	  --   \\
43	&	7	&	2	&	 C$_7$H$_2^{2+}$        	&	  --    	&	0.1	&	0.3	&	0.52	&	0.21	&	 0.26 \\
48.5	&	8	&	2	&	 C$_8$H$^{2+}$         	&	 --     	&	  --    	&	0.39	&	0.55	&	0.25	&	 0.23 \\
49.5	&	8	&	2	&	 C$_8$H$_3^{2+}$        	&	 --     	&	0.06	&	0.65	&	1.6	&	1.4	&	 1.0 \\
50.5	&	8	&	2	&	 C$_8$H$_5^{2+}$        	&	  --    	&	0.06	&	1.13	&	1.3	&	2.6	&	 0.39 \\
51	&	8	&	2	&	 C$_8$H$_6^{2+}$        	&	   --   	&	  --    	&	0.96	&	 --    	&	 --    	&	 --    \\
51.5	&	8	&	2	&	 C$_8$H$_7^{2+}$        	&	  --    	&	0.13	&	0.65	&	0.85	&	0.60	&	 0.41 \\
54.5	&	9	&	2	&	 C$_9$H$^{2+}$         	&	  --    	&	   --   	&	  --    	&	  --   	&	  --   	&	 0.10 \\
55	&	9	&	2	&	 C$_9$H$_2^{2+}$        	&	  --    	&	  --    	&	  --    	&	0.30	&	0.29	&	 0.25 \\
55.5	&	9	&	2	&	 C$_9$H$_3^{2+}$        	&	  --    	&	  --    	&	 --     	&	  --   	&	  --   	&	 0.17 \\
56	&	9	&	2	&	 C$_9$H$_4^{2+}$        	&	  --    	&	   --   	&	  --    	&	   --  	&	0.13	&	  --   \\
59.5	&	9	&	2	&	 C$_9$H$_{10}^{2+}$       	&	    --  	&	   --   	&	0.15	&	  --   	&	  --   	&	     \\
60.5	&	10	&	2	&	 C$_{10}$H$^{2+}$        	&	   --   	&	   --   	&	0.5	&	0.72	&	0.51	&	 0.70 \\
61.5	&	10	&	2	&	 C$_{10}$H$_3^{2+}$       	&	   --   	&	0.01	&	1.1	&	0.33	&	1.1	&	 0.94 \\
62.5	&	10	&	2	&	 C$_{10}$H$_5^{2+}$       	&	   --   	&	0.11	&	0.66	&	1.6	&	0.98	&	 2.0 \\
63	&	10	&	2	&	 C$_{10}$H$_6^{2+}$       	&	  --    	&	  --    	&	0.18	&	 --    	&	  --   	&	 --    \\
63.5	&	10	&	2	&	 C$_{10}$H$_7^{2+}$       	&	   --   	&	0.11	&	0.23	&	1.1	&	0.54	&	 0.53 \\
64	&	10	&	2	&	 C$_{10}$H$_8^{2+}$       	&	0.35	&	0.31	&	0.84	&	 --    	&	  --   	&	  --   \\
64.5	&	10	&	2	&	 $^{13}$CC$_9$H$_8^{2+}$     	&	0.05	&	0.03	&	0.19	&	 --    	&	  --   	&	 --    \\
69	&	11	&	2	&	 C$_{11}$H$_6^{2+}$       	&	 --     	&	  --    	&	 --     	&	  --   	&	0.20	&	   --  \\
69.5	&	11	&	2	&	 C$_{11}$H$_7^{2+}$       	&	   --   	&	 --     	&	 --     	&	  --   	&	0.57	&	  --   \\
70	&	11	&	2	&	 C$_{11}$H$_8^{2+}$       	&	   --   	&	  --    	&	  --    	&	   --  	&	0.16	&	  --   \\
\bottomrule
\end{tabular}
\end{table*}

\begin{table*}
\centering
\label{my-label}
\begin{tabular}{@{}lccccccccc@{}}
\multicolumn{10}{c}{\textbf{Table \textbf{A1}}. Continued.} \\
\toprule
\multirow{2}{*}{m/z} & \multirow{2}{*}{C$_n$} & \multirow{2}{*}{Charge} & \multirow{2}{*}{Attribution} & \multicolumn{3}{c}{Naphthalene} & Anthracene  & 2-Methyl-Anthracene  & Pyrene  \\ 
 &  &  &  & 275 eV & 310 eV & 2500 eV & 2500 eV & 2500 eV & 2500 eV \\ \midrule
72.5	&	12	&	2	&	 C$_{12}$H$^{2+}$        	&	   --   	&	    --  	&	  --    	&	 --    	&	0.17	&	     \\
73.5	&	12	&	2	&	 C$_{12}$H$_3^{2+}$       	&	   --   	&	  --    	&	 --     	&	0.37	&	0.53	&	 1.3 \\
74.5	&	12	&	2	&	 C$_{12}$H$_5^{2+}$       	&	  --    	&	  --    	&	 --     	&	0.60	&	0.32	&	 1.4 \\
75.5	&	12	&	2	&	 C$_{12}$H$_7^{2+}$       	&	 --     	&	  --    	&	  --    	&	0.18	&	0.50	&	 0.46 \\
76	&	12	&	2	&	 C$_{12}$H$_8^{2+}$       	&	  --    	&	  --    	&	 --     	&	1.0	&	   --  	&	   --  \\
76.5	&	12	&	2	&	 C$_{12}$H$_9^{2+}$       	&	  --    	&	  --    	&	 --     	&	0.24	&	0.29	&	 0.18 \\
81	&	13	&	2	&	 C$_{13}$H$_6^{2+}$       	&	   --   	&	   --   	&	  --    	&	  --   	&	0.26	&	   --  \\
81.5	&	13	&	2	&	 C$_{13}$H$_7^{2+}$       	&	  --    	&	 --     	&	  --    	&	  --   	&	1.0	&	  --   \\
82	&	13	&	2	&	 C$_{13}$H$_8^{2+}$       	&	  --    	&	  --    	&	 --     	&	  --   	&	1.1	&	   --  \\
82.5	&	13	&	2	&	 C$_{13}$H$_9^{2+}$       	&	  --    	&	 --     	&	 --     	&	  --   	&	1.7	&	   --  \\
83	&	13	&	2	&	 C$_{13}$H$_{10}^{2+}$      	&	  --    	&	   --   	&	  --    	&	 --    	&	0.91	&	  --   \\
83.5	&	13	&	2	&	 C$_{13}$H$_{11}^{2+}$      	&	  --    	&	  --    	&	 --     	&	  --   	&	0.15	&	   --  \\
84.5	&	14	&	2	&	 C$_{14}$H$^{2+}$        	&	   --   	&	   --   	&	  --    	&	0.22	&	0.25	&	 0.31 \\
85.5	&	14	&	2	&	 C$_{14}$H$_3^{2+}$       	&	  --    	&	 --     	&	  --    	&	0.36	&	0.51	&	 0.77 \\
86.5	&	14	&	2	&	 C$_{14}$H$_5^{2+}$       	&	  --    	&	 --     	&	  --   	&	0.49	&	0.53	&	 0.85 \\
87	&	14	&	2	&	 C$_{14}$H$_6^{2+}$       	&	  --    	&	   --   	&	  --    	&	  --   	&	  --   	&	 1.6 \\
87.5	&	14	&	2	&	 C$_{14}$H$_7^{2+}$       	&	   --   	&	  --    	&	  --    	&	0.34	&	0.50	&	 1.2 \\
88	&	14	&	2	&	 C$_{14}$H$_8^{2+}$       	&	   --   	&	  --    	&	 --     	&	0.70	&	  --   	&	 1.8 \\
88.5	&	14	&	2	&	 C$_{14}$H$_9^{2+}$       	&	   --   	&	  --    	&	  --    	&	0.22	&	0.28	&	 0.58 \\
89	&	14	&	2	&	 C$_{14}$H$_{10}^{2+}$      	&	   --   	&	  --    	&	 --     	&	1.1	&	  --   	&	 --    \\
89.5	&	14	&	2	&	 $^{13}$CC$_{13}$H$_{10}^{2+}$   	&	  --    	&	  --    	&	  --    	&	0.21	&	   --  	&	 --    \\
93.5	&	15	&	2	&	 C$_{15}$H$_7^{2+}$       	&	  --    	&	   --   	&	  --    	&	 --    	&	0.53	&	 --    \\
94	&	15	&	2	&	 C$_{15}$H$_8^{2+}$       	&	   --   	&	   --   	&	 --     	&	  --   	&	1.2	&	  --   \\
94.5	&	15	&	2	&	 C$_{15}$H$_9^{2+}$       	&	  --    	&	  --    	&	 --     	&	 --    	&	1.5	&	  --   \\
95	&	15	&	2	&	 C$_{15}$H$_{10}^{2+}$      	&	  --   	&	  --    	&	  --    	&	  --   	&	1.0	&	 --    \\
95.5	&	15	&	2	&	 C$_{15}$H$_{11}^{2+}$      	&	   --   	&	   --   	&	  --    	&	  --   	&	0.70	&	  --   \\
96	&	15	&	2	&	 C$_{15}$H$_{12}^{2+}$      	&	  --    	&	   --   	&	 --     	&	  --   	&	1.0	&	  --   \\
96.5	&	15	&	2	&	 $^{13}$CC$_{14}$H$_{12}^{2+}$   	&	  --    	&	    -- 	&	   --   	&	  --   	&	0.17	&	  --   \\
97.5	&	16	&	2	&	 C$_{16}$H$_3^{2+}$       	&	   --   	&	  --    	&	 --     	&	   --  	&	  --   	&	 0.42 \\
98.5	&	16	&	2	&	 C$_{16}$H$_5^{2+}$       	&	  --    	&	 --     	&	 --     	&	  --   	&	   --  	&	 0.52 \\
99	&	16	&	2	&	 C$_{16}$H$_6^{2+}$       	&	  --    	&	  --    	&	 --     	&	  --   	&	 --    	&	 0.91 \\
99.5	&	16	&	2	&	 C$_{16}$H$_7^{2+}$       	&	  --    	&	 --     	&	 --     	&	  --   	&	  --   	&	 1.2 \\
100	&	16	&	2	&	 C$_{16}$H$_8^{2+}$ 	&	   --   	&	   --   	&	   --   	&	  --   	&	  --   	&	 4.9 \\
100.5	&	16	&	2	&	 C$_{16}$H$_9^{2+}$       	&	  --   	&	  --    	&	   --   	&	  --   	&	  --   	&	 1.5 \\
101	&	16	&	2	&	 C$_{16}$H$_{10}^{2+}$      	&	  --    	&	  --    	&	  --    	&	  --   	&	  --   	&	 2.9 \\
101.5	&	16	&	2	&	 $^{13}$CC$_{15}$H$_{10}^{2+}$   	&	  --    	&	   --   	&	  --    	&	  --   	&	 --    	&	 0.49 \\
42.7	&	10	&	3	&	 C$_{10}$H$_8^{3+}$      	&	0.02	&	0.01	&	0.47	&	  --   	&	  --   	&	  --   \\
59	&	14	&	3	&	 C$_{14}$H$_9^{3+}$      	&	  --    	&	 --     	&	 --      	&	0.20	&	  --   	&	 --    \\
59.3	&	14	&	3	&	 C$_{14}$H$_{10}^{3+}$     	&	  --    	&	   --   	&	  --    	&	0.29	&	   --  	&	    -- \\
66	&	16	&	3	&	 C$_{16}$H$_6^{3+}$      	&	   --   	&	  --    	&	 --    	&	  --   	&	  --   	&	 0.06 \\
66.3	&	16	&	3	&	 C$_{16}$H$_7^{3+}$      	&	 --     	&	   --   	&	  --    	&	   --  	&	  --   	&	 0.14 \\
66.6	&	16	&	3	&	 C$_{16}$H$_8^{3+}$      	&	  --    	&	  --    	&	 --     	&	  --   	&	 --    	&	 0.56 \\
67	&	16	&	3	&	 C$_{16}$H$_9^{3+}$      	&	  --    	&	  --    	&	  --    	&	  --   	&	  --   	&	 0.55 \\
67.3	&	16	&	3	&	 C$_{16}$H$_{10}^{3+}$     	&	 --     	&	 --     	&	  --    	&	  --   	&	   --  	&	 0.65 \\
67.5	&	16	&	3	&	 $^{13}$CC$_{15}$H$_{10}^{3+}$  	&	  --    	&	 --     	&	 --     	&	   --  	&	   --  	&	 0.12 \\
\bottomrule
\end{tabular}
\end{table*}

\bsp	
\label{lastpage}
\end{document}